\documentclass[twocolumn,amsmath,amsfonts,amssymb,aps,prx,preprintnumbers,superscriptaddress]{revtex4}

\usepackage[utf8]{inputenc}
\usepackage[T1]{fontenc}
\usepackage{lmodern}
\usepackage{graphicx}
\usepackage{dcolumn}
\usepackage{bm}
\usepackage{textcomp}

\usepackage{ifpdf}
\usepackage{physics}
\usepackage{float}
\usepackage[squaren,Gray]{SIunits}
\usepackage{color}
\definecolor{red}{rgb}{1,0,0}
\definecolor{blue}{rgb}{0,0,1}
\definecolor{darkred}{rgb}{0.6,0,0}
\definecolor{darkblue}{rgb}{0,0,0.6}
\definecolor{darkgreen}{rgb}{0,0.5,0}
\definecolor{grey}{rgb}{0.5,0.5,0.5}

\ifpdf
\usepackage{epstopdf}
\usepackage[pdftex,unicode,pdfstartview={FitH},pdfborder={0 0 0}]{hyperref}
\usepackage{hypcap}
\else
\usepackage[hypertex]{hyperref}
\fi
\hypersetup{
    bookmarksnumbered = true,
    colorlinks = true, linkcolor = darkblue,
    citecolor = darkblue, filecolor = darkblue,
    menucolor = darkblue, urlcolor = darkblue
}

\newcolumntype{R}{>{$\displaystyle}r<{$}}
\newcolumntype{C}{>{$\displaystyle}c<{$}}

\begin{document}

\title{Magnetic Correlation Spectroscopy in CrSBr}

\author{Lukas Krelle}
\affiliation{Institute for Condensed Matter Physics, TU Darmstadt, Hochschulstraße 6-8, D-64289 Darmstadt, Germany}

\author{Ryan Tan}
\affiliation{Institute for Condensed Matter Physics, TU Darmstadt, Hochschulstraße 6-8, D-64289 Darmstadt, Germany}

\author{Daria Markina}
\affiliation{Institute for Condensed Matter Physics, TU Darmstadt, Hochschulstraße 6-8, D-64289 Darmstadt, Germany}

\author{Priyanka Mondal}
\affiliation{Institute for Condensed Matter Physics, TU Darmstadt, Hochschulstraße 6-8, D-64289 Darmstadt, Germany}

\author{Kseniia Mosina}
\affiliation{Department of Inorganic Chemistry, University of Chemistry and Technology Prague, Technicka 5, 166 28 Prague 6, Czech Republic}

\author{Kevin Hagmann}
\affiliation{Institute for Condensed Matter Physics, TU Darmstadt, Hochschulstraße 6-8, D-64289 Darmstadt, Germany}

\author{Regine von Klitzing}
\affiliation{Institute for Condensed Matter Physics, TU Darmstadt, Hochschulstraße 6-8, D-64289 Darmstadt, Germany}

\author{Kenji Watanabe}
\affiliation{Research Center for Electronic and Optical Materials, National Institute for Materials Science, 1-1 Namiki, Tsukuba 305-0044, Japan}

\author{Takashi Taniguchi}
\affiliation{Research Center for Materials Nanoarchitectonics, National Institute for Materials Science,  1-1 Namiki, Tsukuba 305-0044, Japan}

\author{Zdenek Sofer}
\affiliation{Department of Inorganic Chemistry, University of Chemistry and Technology Prague, Technicka 5, 166 28 Prague 6, Czech Republic}

\author{Bernhard Urbaszek}
\email{bernhard.urbaszek@pkm.tu-darmstadt.de}
\affiliation{Institute for Condensed Matter Physics, TU Darmstadt, Hochschulstraße 6-8, D-64289 Darmstadt, Germany}

\date{\today}

\begin{abstract}
\noindent CrSBr is an air-stable magnetic van der Waals semiconductor with strong magnetic anisotropy, where the  interaction of excitons with the magnetic order enables the optical identification of different magnetic phases. 
Here, we study the magnetic anisotropy of multi-layer CrSBr inside a three-axis vector magnet and correlate magnetic order and optical transitions in emission and absorption. We identify layer by layer switching of the magnetization through drastic changes of the optical emission and absorption energy and strength as a function of the applied magnetic field. We correlate optical transitions in reflection spectra with photoluminescence (PL) emission using a transfer-matrix analysis and find that ferromagnetic and antiferromagnetic order between layers can coexist in the same crystal. 
In the multi-peak PL emission the intensity of energetically lower lying transitions reduces monotonously with increasing field strength whereas energetically higher lying transitions around the bright exciton $X_{B}$ brighten close to the saturation field. Using this contrasting behavior we can therefore correlate transitions with each other.

\end{abstract}

\maketitle

\section{Introduction}
\noindent Due to a plethora of tunable magnetic intra- and interlayer interactions \cite{gibertini2019magnetic} layered magnetic materials have motivated research towards applications in data storage \cite{song2018giant, wang2018very} and spintronics \cite{ahn20202d, mi2023two}. However, many layered magnetic semiconductors like the chromium trihalides and related materials suffer from instability under ambient conditions \cite{liu2020environmental, galbiati2020very, gish2021ambient, shcherbakov2018raman}. The semiconducting layered antiferromagnet CrSBr overcomes this problem as a stable material in ambient conditions \cite{ziebel2024crsbr,ye2022layer}. It is well suited for fundamental studies of magnons \cite{bae2022exciton, diederich2023tunable, diederich2024exciton}, exciton-phonon coupling \cite{lin2024strong, mondal2024raman,sahu2025resonanceramanscatteringanomalous}, exciton-photon coupling \cite{dirnberger2023magneto, wang2023magnetically,li20242d,nessi2024magnetic,han2025exciton} as well as applications in magnetic devices exhibiting large negative magnetoresistance \cite{boix2022probing, telford2020layered}, tunable properties with magnetic fields \cite{wilson2021interlayer} and electrostatic doping \cite{tabataba2024doping}. CrSBr crystals are strongly anisotropic which results in strong dependence on the crystallographic axes also for magnetic, optical and electronic properties \cite{klein2023bulk, wilson2021interlayer, yang2021triaxial}. \\
\noindent The strong coupling between excitons and the magnetic order in CrSBr enables the spectroscopic identification of different magnetic phases \cite{wilson2021interlayer, tabataba2024doping,liebich2025controlling} as well as spatial domains \cite{tabataba2024doping}. The spectroscopic signature of excitonic transitions is not limited to the topmost layer but also reveals information about deeper lying layers \cite{shao2025magnetically}, which is potentially an advantage compared to successful scanning probe techniques \cite{tschudin2024imaging, rizzo2022visualizing}. Up to now, research has mainly focused on understanding the magnetic and excitonic properties of bulk systems and single to 4 layer systems. However, the study of multilayer systems, which exhibit complex emission properties without the influence of strong exciton-photon coupling, has remained elusive. \\
\noindent In this work, we study the anisotropic magnetic properties of CrSBr inside a vector magnet at cryogenic temperatures. We use correlated spectral changes occurring with application of a magnetic field to investigate the origin and connections of several emission lines present in multilayer samples of CrSBr. To this end, we fabricated two multilayer samples, one (14 layers, thickness determined by atomic force microscopy 11.3 $\pm$ 0.2 nm) encapsulated in hBN, another one as-exfoliated (10 layers, thickness determined by atomic force microscopy 8.2 $\pm$ 0.3 nm). We performed magnetic field dependent photoluminescence (PL) and differential reflectance contrast (DR/R) measurements at T = 4.7 K and carefully track emission energy and intensity to conclude on potential emission origins. Our measurements show, that the ferromagnetic state is reached gradually throughout the crystal as different layers switch from the antiferromagnetic to the ferromagnetic order at different fields. We find phases close to the saturation field, that show a superposition of spectral signatures of both the ferromagnetic (FM) and antiferromagnetic (AFM) state.

\begin{figure*}[t!]
\includegraphics[scale=1]{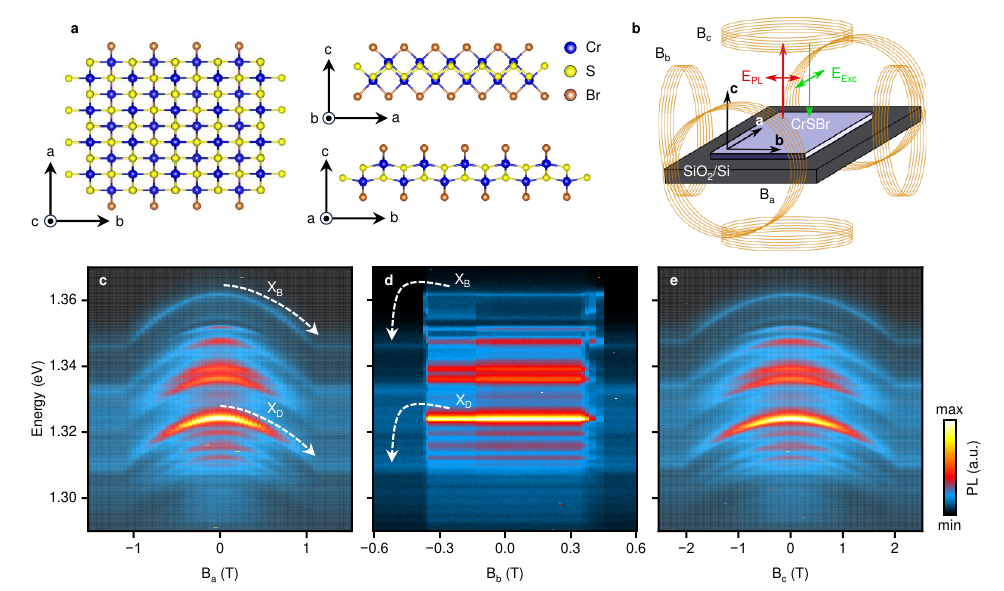}
\caption{\textbf{Experiments on CrSBr in three-axis vector magnet} a: CrSBr crystal structure. Crystallographic axes a,b,c are indicated by arrows. b: Sketch of a CrSBr sample inside the vector magnet and the orientation of Laser and PL polarization with respect to crystallographic axes. c-e: Exemplary PL magnetic field sweeps of the encapsulated sample for the same spatial spot along the crystal a-,b- and c-axis respectively displaying the magnetic anisotropy, where B$_{a}$ $||$ a, B$_{b}$ $||$ b and B$_{c}$ $||$ c. Dashed white arrows are a guide for the eye for evolution of $X_D$ (at 1.324 eV at B=0)  and $X_B$ (at 1.362 eV at B=0) transitions.} 
\label{fig:CrystalandBaxes}
\end{figure*}

\begin{figure*}[t!]
\includegraphics[scale=1]{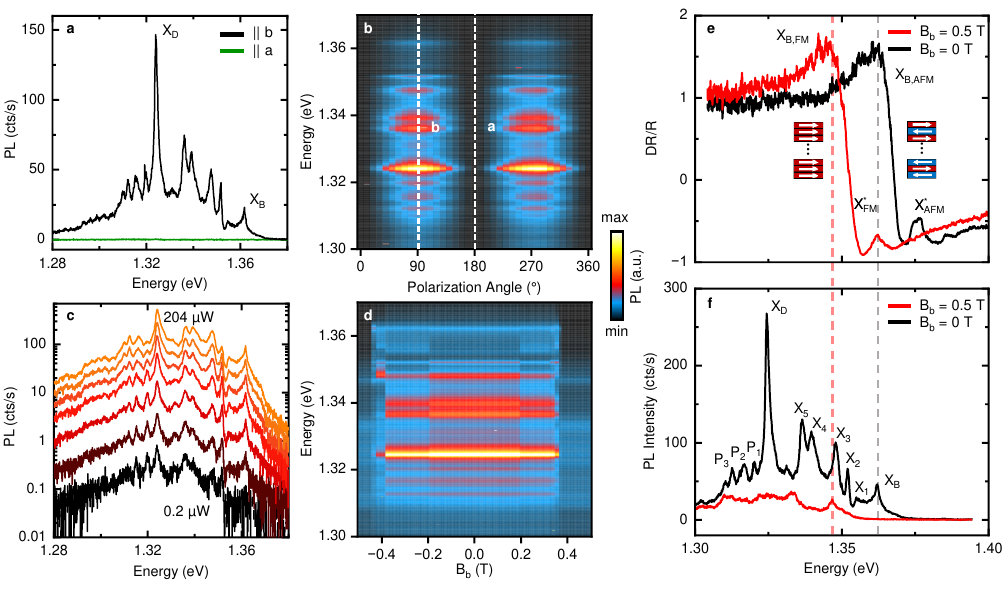}
\caption{\textbf{Photoluminescence and differential reflectance contrast of encapsulated 14 layer CrSBr at T = 4.7 K.} a: PL spectra at B = 0 T with detection polarization aligned with the crystal b- (black solid line) and a-axis (green solid line). b: Full polarization dependence of the emission. c: Power dependence of the PL emission at B = 0 T. d: PL magnetic field sweeps along the crystal b-axis for an encapsulated (14 L) sample. Magnetic field step size is 10 mT, sweep direction from positive to negative fields. e-f: DR/R and PL spectra of the 14 layer sample for the AFM-state (black) and the FM-state (red). Emissions indicated as in main text. Dashed gray and red lines indicate the energy of X$_{B}$ in the AFM and FM state respectively.} 
\label{fig:general_specs}
\end{figure*}

\section{Results and Discussion}

\noindent We performed cryogenic magneto-optical spectroscopy inside a three-axis vector magnet, in which the crystallographic axes are aligned with the magnetic field axes. Figures \ref{fig:CrystalandBaxes} \textbf{a} and \textbf{b} display the crystal structure of CrSBr and the experimental configuration respectively. Unless noted differently, we focus on the hBN encapsulated 14 layer sample. We performed PL measurements at T = 4.7 K with an excitation energy of 1.959 eV, above the estimated band gap of CrSBr \cite{wilson2021interlayer, klein2023bulk}. First, we study the magnetic anisotropy of the material, which is clearly visible in the magnetic field sweeps along the crystal axes shown in Figure \ref{fig:CrystalandBaxes} \textbf{c}-\textbf{e} for the same sample spot. At zero magnetic field, below the critical temperature, the layers show  PL spectra associated with antiferromagnetic order, as spins in adjacent layers point in the opposite direction along the crystal b-axis \cite{ye2022layer, lee2021magnetic, goser1990magnetic}. The application of a magnetic field rotates the magnetization such that beyond a certain saturation field, all layers are magnetized along the same direction (i.e. parallel to the applied field) corresponding to ferromagnetic order. We trace the change from AFM to FM order through changes in the PL energy \cite{wilson2021interlayer}. For magnetic fields applied along the magnetic hard (c-axis) and the magnetic intermediate axis (a-axis), the spins rotate gradually while for magnetic fields applied along the magnetic easy axis (b-axis), the spins flip abruptly \cite{wilson2021interlayer}. The saturation fields along the different axis for the data shown are, respectively $B_c^s = \pm 2.05 \pm 0.05$ T and $B_a^s = \pm 1.10 \pm 0.05$ T. Due to hysteresis effects the saturation fields along b depend on the sweep direction with $B_b^s = +0.45 \pm 0.01$ T and $B_b^s = -0.37 \pm 0.01$ T. The energy shift for sweeps along the a-axis follows a stronger curvature with applied field compared to sweeps along the c-axis \cite{PhysRevB.111.075107}.

\begin{figure*}[t]
\includegraphics[scale=1]{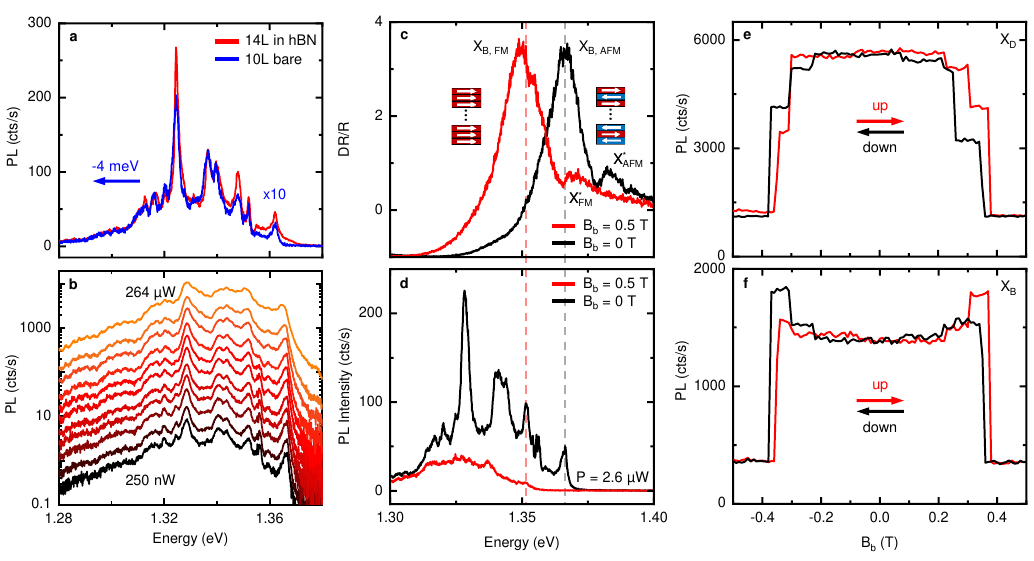}
\caption{\textbf{Characteristics of the unencapsulated 10L sample.} a: PL comparison of the two samples. Excitation power was 77.6 $\mu$W  for the encapsulated sample was and 250 nW for the unencapsulated sample. The spectrum of the unencapsulated sample was redshifted by 4 meV for better comparison. b: Powerdependence of the PL emission. c-d: DR/R and PL for the AFM (black) and FM (red) phase. Dashed gray and red lines indicate the energy of X$_{B}$ in the AFM and FM state respectively. e-f: Maximum intensity plots of $X_{D}$ (e) and $X_{B}$ (f) for the unencapsulated 10L sample for up-sweeping (down-sweeping) magnetic field in red (black). } 
\label{fig:Sup3}
\end{figure*}

The continuous energy shift of the emission peaks for sweeps along the magnetic hard and intermediate axis allows the assignment of the emission peaks in the AFM and FM state, which we aim to identify. For comparison, the abrupt jumps at $B_b^s$ make peak assignments between optical emission in the AFM and FM states more challenging. \\
In our PL measurements, the linear excitation polarization is aligned with the crystal a-axis, while the detection polarization is aligned with the b-axis (see Figure \ref{fig:CrystalandBaxes} \textbf{b}). The crystal axes were determined via the maximum and minimum PL intensity for the b-axis and a-axis respectively. Figure \ref{fig:general_specs} \textbf{a} and \textbf{b} show the polarization dependence of the PL emission of encapsulated 14 layer CrSBr at B = 0 T. Due to the strong anisotropy of the material, the PL emission is highly linearly polarized exhibiting maximum intensity along the b-axis and about a factor of 200 weaker intensity along the a-axis. Both samples show qualitatively similar PL spectra with a multitude of different emission peaks with narrow linewidths (of the order of meV) as shown in Figure \ref{fig:general_specs} \textbf{a} for encapsulated 14 layer and in \ref{fig:Sup3} \textbf{a} for unencapsulated 10 layer CrSBr. Importantly, due to the small sample thickness, the rich spectra cannot stem from polaritonic states, as reported for bulk CrSBr layers \cite{dirnberger2023magneto, wang2023magnetically}, but from different excitonic species and their phonon replica, which we aim to identify further. \\

\noindent In Figure \ref{fig:general_specs} \textbf{a}, we mark two transitions, that will be of significance in our discussion. The transition $X_D$, which is the strongest in PL and the transition $X_B$, which is visible in PL and is the strongest transition in terms of oscillator strength in DR/R, see below. Observing strong PL from transitions that have low oscillator strength resembles the rich PL spectra of WSe$_2$ monolayers with several exciton species \cite{wang2017plane,he2020valley,shree2021guide}.\\

\noindent We performed excitation power dependent measurements, displayed in Figure \ref{fig:general_specs} \textbf{c}, to ensure, that we capture all details of the complex spectra. Since the spectra in the investigated power range experience only slight changes in their relative intensities, we continue with a moderate excitation power of about 60 $\mu$W. In the unencapsulated sample, the changes from low to high power are more pronounced as shown in Figure \ref{fig:Sup3} \textbf{b}. Here we focus on a low excitation power regime, in which both samples exhibit similar emission properties as highlighted in the direct comparison in Figure \ref{fig:Sup3} \textbf{a}. We note, that the emission peaks in the unencapsulated sample are blueshifted by 4 meV with respect to the encapsulated sample, most likely due dielectric screening \cite{raja2017coulomb, stier2016probing}, while relative energy splittings within the spectra remain unchanged.

\begin{figure}[h]
\includegraphics[scale=1]{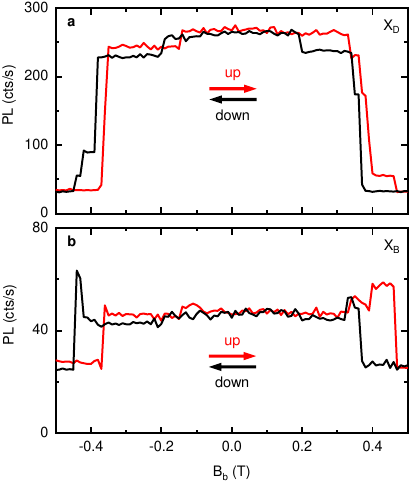}
\caption{\textbf{Hysteresis of PL intensities in magnetic fields along $B_b$.} a-b: Maximum intensity plots of $X_{D}$ (a) and $X_{B}$ (b) for the encapsulated 14L sample for up-sweeping (down-sweeping) magnetic field in red (black).} 
\label{fig:Hysteresis}
\end{figure}

\begin{figure*}[t!]
\includegraphics[scale=1]{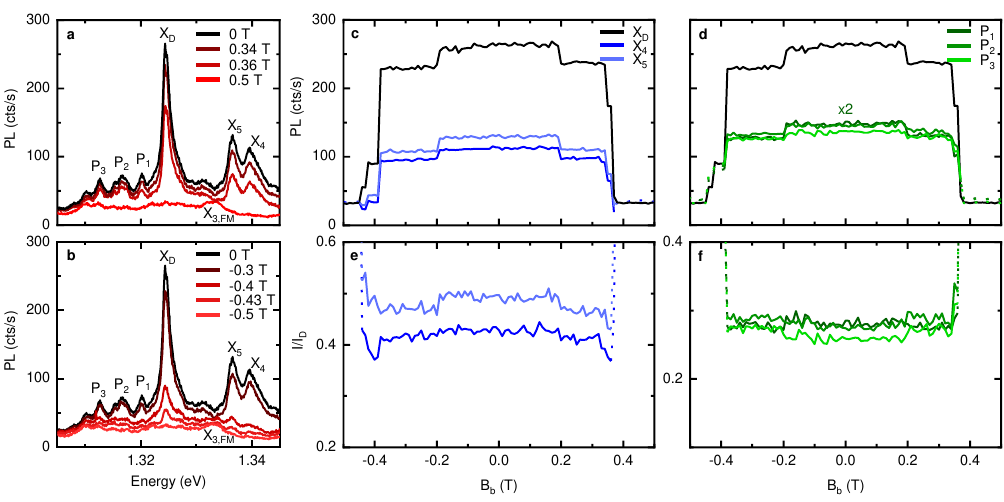}
\caption{\textbf{Correlated emissions in magnetic fields.} a: PL emission of the encapsulated CrSBr flake for selected magnetic field strengths $B_{b}$ along the crystal b-axis and spectral range displaying similar and consistent reduction of emission intensity with magnetic field. b: same as (a), but for negative magnetic fields. Gray arrow is a guide for the eye. c-d: Maximum intensity of the emissions marked in (a) for the magnetic field sweep in Figure \ref{fig:general_specs} d. e-f: Maximum emission intensity in c-d normalized by the intensity of $X_{D}$. Pointed parts of the graphs indicate regions with large error bars.} 
\label{fig:Intensity_Tracking1}
\end{figure*}

\begin{figure*}[t]
\includegraphics[scale=1]{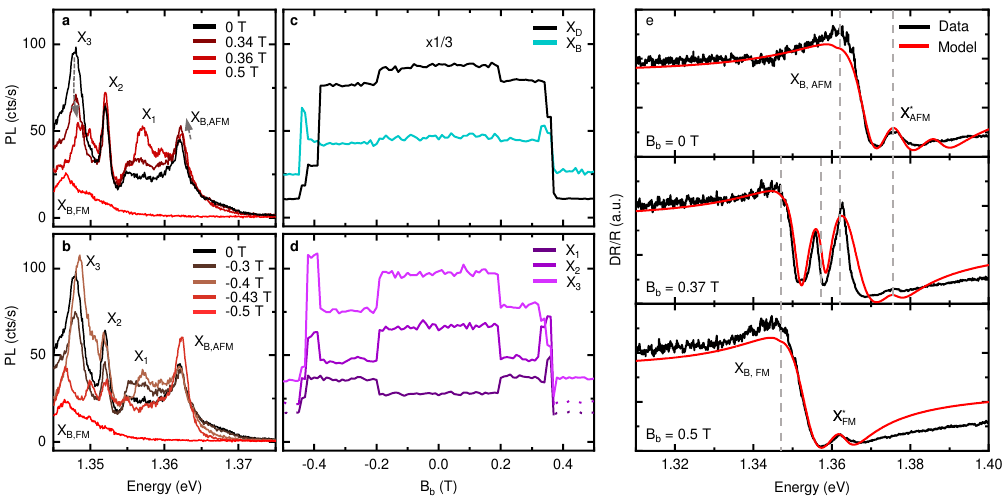}
\caption{\textbf{Anti-correlated emissions and superimposed magnetic phases.} a: PL emission for selected magnetic field strengths and spectral range along the crystal b-axis showing unexpected brightening of emissions. b: same as (a), but for negative magnetic fields. Grey arrows are guides for the eye. c: Maximum intensity of the $X_{B}$ and $X_{D}$ emissions for magnetic field sweep in Figure \ref{fig:general_specs} d. X$_{D}$ (X$_{B}$) reduces in intensity by 9.5 $\pm$ 2 $\%$ (4.1 $\pm$ 4) $\%$ for $B_{b}$ > 0.18 T and by 12.3 $\pm$ 2 $\%$ (8.1 $\pm$ 4.2) $\%$ for $B_{b}$ < -0.2 T. d: Maximum intensity of the remaining emissions marked in (a). Pointed parts of the graphs indicate regions with large error bars. e: Transfer-matrix-analysis of DR/R measurements for selected magnetic fields. Dashed grey lines indicate the energies of the respective oscillators.} 
\label{fig:Intensity_Tracking2}
\end{figure*}

\noindent Theoretical calculations predict two conduction bands resulting in two possible transitions to the valence band of which one is parity allowed and one is parity forbidden \cite{wilson2021interlayer, klein2023bulk}. The notion of parity allowed and forbidden transitions only strictly holds for transitions at very specific points in k-space, which is not fully applicable to the more extended exciton states in CrSBr \cite{bianchi2023paramagnetic}. As proposed in other work \cite{lin2024probing, wilson2021interlayer}, we refer to the emission around 1.362 eV ($X_{B}$) to the parity allowed transition.  This is in accordance with DR/R measurements displayed in Figure \ref{fig:general_specs} \textbf{e}, which reveal an optical resonance with large oscillator strength at the same energy. This resonance is accompanied by a much weaker resonance roughly 14 meV above this transition (see transition labeled $X^*$ in Figure \ref{fig:Intensity_Tracking2} \textbf{e}). Klein et al. \cite{klein2023bulk} suggest that this resonance might have contributions from the same conduction band along the $\Gamma-X$ direction. 
In the PL measurements shown the emission of this transition is absent due to the low excitation power. For the assignment of the so-called parity forbidden transition $X_{D}$, we refer to work by Lin et al. \cite{lin2024strong, lin2024probing}. This corresponds to the brightest transition in PL emission located around 1.324 eV, 38 meV below $X_{B}$ in agreement with previous reports and theoretical predictions for the splitting of the exciton energies \cite{lin2024probing, wilson2021interlayer}. We note, that the magnitude of the bandgap energy and the origin and magnitude of the splitting between $X_B$ and $X_D$ is discussed in several studies \cite{klein2023bulk, komar2024colossal, smolenski2025large, qian2023anisotropic, datta2024magnon, shao2025magnetically}. Our work is adding contrasting magnetic field effects to the observed distinctions between the two transitions. Importantly, we do not observe any signature of $X_{D}$ in DR/R in agreement with its parity forbidden character i.e. predicted weak oscillator strength. Between $X_{B}$ and $X_{D}$ we find several emission peaks, that we label $X_{1-5}$ as well as three emissions below $X_{D}$, which we label $P_{1-3}$. \\

\noindent To gain further insight into the origin of the emission peaks $X_{1-5}$ and $P_{1-3}$, we performed magnetic field sweeps along the magnetic easy axis (for details see methods). We recorded a full hysteresis loop starting in the FM state at negative magnetic field $B_{b}$ and focus on the down-sweep direction of the measurement loop shown in Figure \ref{fig:general_specs} \textbf{d}. Figure \ref{fig:Hysteresis} displays the maximum intensity of $X_{D}$ and $X_{B}$ for the whole magnetic field sweep including strong hysteresis effects. For both samples, we observe several abrupt changes in the emission intensities at distinctly different magnetic fields. Hence, the magnetization of the sample changes in discrete steps, i.e. different layers in the sample change their spin orientation individually. We note a certain analogy to layer by layer switching in Fe/MgO (001) superlattices \cite{moubah2016discrete}, albeit we use here simple PL and DR/R measurements to monitor the magnetism in CrSBr. Importantly, different PL emission peaks change at the same magnetic field demonstrating the strong coupling of excitons to the magnetic order of the system. 
The emission changes most drastically, when the magnetic field reaches the saturation field and the remaining layers of the system switch to the FM state. For the down-sweep direction in Figure \ref{fig:general_specs} \textbf{d}, this happens at $B_{b}$ < -0.44 T in the encapsulated sample. In the unencapsulated sample, the switch happens at  $B_{b}$ < -0.37 T. Due to hysteresis, the FM state starts to vanish at a different field $B_{b}$ < 0.38 T in the encapsulated sample and $B_{b}$ < 0.36 T in the unencapsulated sample.
In the FM state, the emission intensity is much weaker than in the antiferromagnetic (AFM) state and the emissions experience a redshift of 16 meV for $X_{B}$ and 14 meV for $X_{D}$. This reduction of intensity is thought to result from the reduced layer confinement of excitons in the FM state \cite{marques2023interplay}. In the AFM state, the antiparallel magnetization of adjacent layers does not allow charge transfer between layers and thus excitons are confined to a layer. However, the parallel magnetization of adjacent layers in the FM state allows charge transfer reducing the electron-hole wavefunction overlap. Strikingly, all emission peaks in the investigated samples are weaker in the FM states while the exact changes during the transition can differ which we will elaborate further. In the FM state, the intensity of $X_{B}$ and $X_{D}$ reduce by approximately $46 \pm 3 \%$ and $88 \pm 1 \%$ respectively. \\
\medskip

\noindent We now turn to a more detailed discussion of the emission changes. Figure \ref{fig:Intensity_Tracking1} \textbf{a} and \textbf{b} display spectra recorded at different values of $B_{b}$ in a spectral range around the strongest peak in PL emission $X_{D}$. With increasing magnetic field strength, all of the emission peaks in this range reduce in intensity until either vanishing (within our detection limit) or experiencing an energy shift at the switch to the FM phase. A moderate reduction in intensity appears at $B_{b} = \pm$ 0.2 T, which likely corresponds to a switch of an individual layer magnetization. 
Figure \ref{fig:Intensity_Tracking1} \textbf{c} (for transitions $X_D$, $X_4$, $X_5$) and \textbf{d} (for transitions $P_1$, $P_2$, $P_3$) highlight the collective change, where we plot the maximum intensity of each of the emissions. We emphasize this further by plotting the intensity ratios of the emissions compared to the emission of $X_{D}$ in Figure \ref{fig:Intensity_Tracking1} \textbf{e} and \textbf{f}. Until the switching to the fully ferromagnetic phase, the intensity ratios stay almost constant. We want to emphasize that in addition, before reaching B$_b^S$ none of these emissions exhibit an energy shift despite the changes in intensity. Due to this correlated behaviour, these emissions might stem from the same band transition. \\
The emissions $P_{1}$, $P_{2}$ and $P_{3}$ are redshifted with respect to $X_{D}$ by 4.2 meV, 7.8 meV and 11.8 meV respectively, i.e. approximately equidistantially spaced by 4 meV. From this one might speculate, that the emissions $P_{1}$, $P_{2}$ and $P_{3}$ are phonon replica of $X_{D}$. Pawbake et al. \cite{pawbake2023raman} report, that the increased size of the unit cell in the AFM phase leads to backfolding of several phonon modes to the $\Gamma$-point compared to the FM phase. Consequently, several acoustic phonon modes in the energy range of 20-40 cm$^{-1}$ (2.5 - 5 meV) could in principle couple to the exciton and create the observed equidistant emissions. In the FM phase, these phonon modes would gain finite quasi-momentum and thus the replica would disappear again. Within the resolution of the observed PL spectra we cannot identify the presence of $P_{1-3}$ in the FM phase. We note, that in the energy range of $P_{1-3}$ phonon replica of $X_{D}$ with the $A^{1}_{g}$ mode have been reported \cite{lin2024strong}. At higher excitation powers, we find signatures of these emissions overlayed with $P_{1-3}$. The attribution of $X_{4}$ and $X_{5}$, centered around 1.340 eV and 1.337 eV respectively, need to be clarified in further studies. At this point we can report that $X_4$ and $X_5$ show an evolution with magnetic field B$_{b}$, that is similar to $X_D$. \\
\medskip

\noindent While the emission peaks between 1.31 eV and 1.345 eV experience very similar emission changes in magnetic fields compared to $X_D$, the emissions between 1.345 eV and 1.37 eV behave very differently. Figure \ref{fig:Intensity_Tracking2} \textbf{a} and \textbf{b} show spectra in this energy range for the same magnetic field values as in Figure \ref{fig:Intensity_Tracking1} \textbf{a} and \textbf{b}. At zero magnetic field, the emission energies of $X_{1-3}$ are located at $E_{1} = 1.355$ eV, $E_{2} = 1.352$ eV and $E_{3} = 1.348$ eV. Most strikingly, the intensity of $X_{B}$ as a function of $B_b$ evolves very differently compared to $X_{D}$, as highlighted in Figure \ref{fig:Intensity_Tracking2} \textbf{c} and comparing Figures \ref{fig:Hysteresis} \textbf{a} and \textbf{b}. As described before, the intensity of $X_{D}$ reduces measurably as individual layers flip spin orientation. In contrast, the intensity of $X_{B}$ reduces less as highlighted in Figure \ref{fig:Intensity_Tracking2} \textbf{c}. Very surprisingly, at magnetic fields close to the transition to the FM state, the intensity of $X_{B}$ increases. Similar to $X_{B}$, also the transitions $X_{1}$, $X_{2}$ and $X_{3}$ experience an increase in intensity close to the saturation field. Additionally, $X_{2}$ and $X_{3}$ initially decrease in intensity with the single layer switch at $B_b=\pm 0.2$ T. In contrast, at $B_b=\pm 0.2$ T the transition $X_{1}$ increases its intensity. Despite the changes, $X_{B}$ and  $X_{2}$ do not shift in energy before the transition to the FM state. However, $X_{1}$ and $X_{3}$ experience a slight blueshift as shown in Figure \ref{fig:Intensity_Tracking2} \textbf{a} and \textbf{b}.\\

\begin{figure*}[t]
\includegraphics[scale=1]{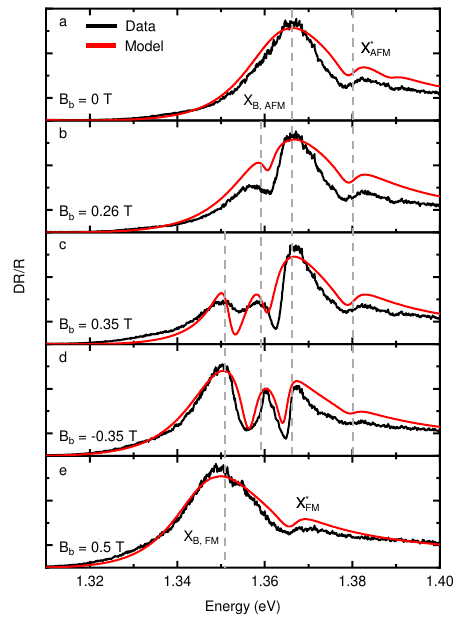}
\caption{Transfer-matrix-analysis of DR/R measurements of the unencapsulated 10L sample for selected magnetic fields. Dashed lines indicate the energies of the respective oscillators.} 
\label{fig:Sup5}
\end{figure*}

\noindent Above B$^{S}_{b}$ the entire crystal is in the FM state. For applied magnetic fields 0 < B$_{applied}$ < B$^{S}_{b}$ the AFM and FM configuration can coexist in the crystal and we observe spectral features of both configurations in PL and DR/R measurements. In PL, the emission peak of e.g. $X_{3, FM}$ is overlayed with the weakened emission peaks of $X_{4}$ and $X_{5}$ in the AFM configuration as highlighted in Figure \ref{fig:Intensity_Tracking1} \textbf{b}. Similarly, the emission peak of $X_{B, FM}$ is overlayed with the emission peak of $X_{3}$ displayed in Figure \ref{fig:Intensity_Tracking2} \textbf{a} and \textbf{b}.
Further insight comes from DR/R measurements at magnetic fields close to the transition displayed in Figure \ref{fig:Intensity_Tracking2} \textbf{e}. In addition to resonance of $X_{B,AFM}$, another resonance appears at 1.347 eV, which we identify as the resonance of $X_{B}$ in the FM state ($X_{B,FM}$). The resonance of $X_{B,FM}$ appears several 10 mT below the saturation field and persists above it. Surprisingly, the measurements reveal an additional resonance at 1.357 eV, redshifted by 5 meV from $X_{B,AFM}$ around 1.362 eV, as shown in Figure \ref{fig:Intensity_Tracking2} \textbf{e}. This resonance appears with the first switch of magnetization and remains present until the system fully transitions into the FM state, where it vanishes. The origin of this resonance remains to be clarified. We use a transfer-matrix-analysis (for details see methods) to reconstruct the measurements and determine the resonance energies of the oscillators present in the different magnetic phases. This analysis is necessary as the resonance energies do not coincide with extrema of the DR/R spectra due to interference effects and the finite thickness of the CrSBr layers. We are able to reproduce the spectra well with our formalism allowing to determine the resonance energies present in the spectrum with an estimated accuracy of 1-2 meV. For consistency, we checked the unencapsulated sample in several spots and it exhibits a similar superposition of the individual resonances as highlighted in Figure \ref{fig:Sup5}.\\

\noindent \textit{In conclusion}, tuning the magnetic field in a three-axis vector magnet gives access to the different spin configurations in multilayer CrSBr. Photoluminescence measurements reveal step-like changes in intensity for all observed emission peaks correlated to a layer by layer switching of the magnetization for fields applied along the magnetic easy axis. Some of these emissions around the parity forbidden transition $X_{D}$ reduce monotonously with increasing magnetic field and do not experience energy shifts until the saturation field along $B_b$. We find several exceptions from this trend between $X_{D}$ and $X_{B}$, that do experience energy shifts and even increase in intensity close to the saturation field. Additionally, we find, that in phases close to the transition to the FM state, excitonic emission from the AFM and FM phases coexists. We show that the comparison between PL and reflectivity measurements gives access to the magnetization configurations within the multilayer.

\section{Methods}

\subsection{Sample fabrication:}
\noindent Bulk CrSBr crystals were fabricated through chemical vapor transport \cite{klein2022control}. Nanometer thin CrSBr and hBN flakes were mechanically exfoliated onto PDMS and transferred onto Si substrates with an 80 nm thick oxide layer. The layer thicknesses were determined using an atomic force microscope (Oxford Instruments Cypher) equipped with AC160 cantilevers (Oxford Instruments). Atomic force microscopy yields CrSBr thickness of 8.2 $\pm$ 0.3 nm (11.3 $\pm$ 0.2 nm) corresponding to 10 (14) layers for the unencapsulated (encapsulated) sample. Both samples were annealed at 200$^{\circ}$ for several hours. 

\subsection{Optical spectroscopy:}
\noindent Optical spectroscopy was carried out in a home-built confocal setup for magneto-optical spectroscopy \cite{shree2021guide}. The sample was placed inside a closed-cycle cryostat (attocube systems, AttoDry 1000XL) equipped with a vector magnet (z-axis: solenoid, maximum field 5 T, x-/y-axis: Split Coil, maximum field 2 T). We used low temperature piezo-positioners (attocube systems, ANPx101 and ANPz102) to position the sample with respect to a low temperature apochromatic objective. PL and DR/R measurements were performed in backscattering geometry at a sample temperature of 4.7 K. The signal was dispersed inside a Czerny-Turner spectrograph (Teledyne Princeton Instruments, SpectraPro HRS-500) and detected by a CCD-camera (Teledyne Princeton Instruments, Pylon BRexcelon 100). For DR/R measurements we used a Tungsten-Halogen lamp (Thorlabs, SLS201L/M) polarized along the crystal b-axis by a nanoparticle-film polarizer and an achromatic half-waveplate. For PL measurements we used a HeNe Laser (Thorlabs, HNL210LB) with its polarization aligned along the crystal a-axis while the emission was detected along the b-axis. PL measurements were performed at excitation powers ranging from 250 $nW$ to 260 $\mu W$. Magnetic field dependent measurements were performed initializing the CrSBr sample in the FM-state, ramping the magnet to -0.6 T (-0.5 T) for the encapsulated 14 layers (unencapsulated, 10 layers) sample, followed by a sweep to 0.6 T (0.5 T) and subsequent inversion of the sweep direction to obtain a full hysteresis.

\subsection{Transfer-matrix analysis}
\noindent For the analysis of differential reflectance contrast measurements, we apply a transfer-matrix formalism \cite{dirnberger2023magneto, wang2023magnetically, robert2018optical}, using a Lorentzian oscillator model for the dielectric constant of CrSBr 
\begin{equation}
    \epsilon(\omega) = \epsilon_{\infty} + \sum_{j} \frac{f_{j}/\hbar^{2}}{\omega_{j}^{2} - \omega^{2} - i\Gamma_{j} \omega}
\end{equation}
where $\omega_{j}$ and $\Gamma_{j}$ denote the oscillator frequency and decay rate and $f_{j}$ denotes the oscillator strength of the $jth$ oscillator. We account for a constant background permittivity $\epsilon_{\infty} = 10$, similar to Wang et al \cite{wang2023magnetically} and use the permittivities $\epsilon_{SiO_{2}} = 2.1$ \cite{malitson1965interspecimen}, $\epsilon_{Si} = 13$ \cite{schinke2015uncertainty} and $\epsilon_{hBN} = 4.6$ \cite{lee2019refractive}.

\vspace{8pt}
\noindent \textbf{Acknowledgements:} \\
We thank Florian Dirnberger for fruitful discussions.\\
K.W. and T.T. acknowledge support from the JSPS KAKENHI (Grant Numbers 21H05233 and 23H02052) , the CREST (JPMJCR24A5), JST and World Premier International Research Center Initiative (WPI), MEXT, Japan. Z.S. was supported by ERC-CZ program (project LL2101) from Ministry of Education Youth and Sports (MEYS) and by the project Advanced Functional Nanorobots (reg. No. CZ.02.1.01/0.0/0.0/15$_{-}$003/0000444 financed by the EFRR).\\

\noindent \textbf{Author Contributions:} 
K.M. and Z.S. grew bulk CrSBr crystals. T.T. and K.W. grew bulk hBN crystals. L.K. fabricated the samples and performed optical spectroscopy with R.T.. P.M., L.K., K.H., R.v.K. performed and analysed AFM measurements. L.K., R.T. and B.U. analyzed the optical spectra. L.K., R.T., P.M., D.M. and B.U. discussed the results. B.U. suggested the experiments and supervised the project. L.K., R.T. and B.U. wrote the manuscript. All authors contributed to the final manuscript.\\

\noindent \textbf{Competing interests}: The authors declare no competing interests.\\

\vspace{8pt}


\newpage

\begin{thebibliography}{54}
\expandafter\ifx\csname natexlab\endcsname\relax\def\natexlab#1{#1}\fi
\expandafter\ifx\csname bibnamefont\endcsname\relax
  \def\bibnamefont#1{#1}\fi
\expandafter\ifx\csname bibfnamefont\endcsname\relax
  \def\bibfnamefont#1{#1}\fi
\expandafter\ifx\csname citenamefont\endcsname\relax
  \def\citenamefont#1{#1}\fi
\expandafter\ifx\csname url\endcsname\relax
  \def\url#1{\texttt{#1}}\fi
\expandafter\ifx\csname urlprefix\endcsname\relax\def\urlprefix{URL }\fi
\providecommand{\bibinfo}[2]{#2}
\providecommand{\eprint}[2][]{\url{#2}}

\bibitem[{\citenamefont{Gibertini et~al.}(2019)\citenamefont{Gibertini,
  Koperski, Morpurgo, and Novoselov}}]{gibertini2019magnetic}
\bibinfo{author}{\bibfnamefont{M.}~\bibnamefont{Gibertini}},
  \bibinfo{author}{\bibfnamefont{M.}~\bibnamefont{Koperski}},
  \bibinfo{author}{\bibfnamefont{A.~F.} \bibnamefont{Morpurgo}},
  \bibnamefont{and} \bibinfo{author}{\bibfnamefont{K.~S.}
  \bibnamefont{Novoselov}}, \bibinfo{journal}{Nature nanotechnology}
  \textbf{\bibinfo{volume}{14}}, \bibinfo{pages}{408} (\bibinfo{year}{2019}).

\bibitem[{\citenamefont{Song et~al.}(2018)\citenamefont{Song, Cai, Tu, Zhang,
  Huang, Wilson, Seyler, Zhu, Taniguchi, Watanabe et~al.}}]{song2018giant}
\bibinfo{author}{\bibfnamefont{T.}~\bibnamefont{Song}},
  \bibinfo{author}{\bibfnamefont{X.}~\bibnamefont{Cai}},
  \bibinfo{author}{\bibfnamefont{M.~W.-Y.} \bibnamefont{Tu}},
  \bibinfo{author}{\bibfnamefont{X.}~\bibnamefont{Zhang}},
  \bibinfo{author}{\bibfnamefont{B.}~\bibnamefont{Huang}},
  \bibinfo{author}{\bibfnamefont{N.~P.} \bibnamefont{Wilson}},
  \bibinfo{author}{\bibfnamefont{K.~L.} \bibnamefont{Seyler}},
  \bibinfo{author}{\bibfnamefont{L.}~\bibnamefont{Zhu}},
  \bibinfo{author}{\bibfnamefont{T.}~\bibnamefont{Taniguchi}},
  \bibinfo{author}{\bibfnamefont{K.}~\bibnamefont{Watanabe}},
  \bibnamefont{et~al.}, \bibinfo{journal}{Science}
  \textbf{\bibinfo{volume}{360}}, \bibinfo{pages}{1214} (\bibinfo{year}{2018}).

\bibitem[{\citenamefont{Wang et~al.}(2018)\citenamefont{Wang,
  Guti{\'e}rrez-Lezama, Ubrig, Kroner, Gibertini, Taniguchi, Watanabe,
  Imamo{\u{g}}lu, Giannini, and Morpurgo}}]{wang2018very}
\bibinfo{author}{\bibfnamefont{Z.}~\bibnamefont{Wang}},
  \bibinfo{author}{\bibfnamefont{I.}~\bibnamefont{Guti{\'e}rrez-Lezama}},
  \bibinfo{author}{\bibfnamefont{N.}~\bibnamefont{Ubrig}},
  \bibinfo{author}{\bibfnamefont{M.}~\bibnamefont{Kroner}},
  \bibinfo{author}{\bibfnamefont{M.}~\bibnamefont{Gibertini}},
  \bibinfo{author}{\bibfnamefont{T.}~\bibnamefont{Taniguchi}},
  \bibinfo{author}{\bibfnamefont{K.}~\bibnamefont{Watanabe}},
  \bibinfo{author}{\bibfnamefont{A.}~\bibnamefont{Imamo{\u{g}}lu}},
  \bibinfo{author}{\bibfnamefont{E.}~\bibnamefont{Giannini}}, \bibnamefont{and}
  \bibinfo{author}{\bibfnamefont{A.~F.} \bibnamefont{Morpurgo}},
  \bibinfo{journal}{Nature communications} \textbf{\bibinfo{volume}{9}},
  \bibinfo{pages}{2516} (\bibinfo{year}{2018}).

\bibitem[{\citenamefont{Ahn}(2020)}]{ahn20202d}
\bibinfo{author}{\bibfnamefont{E.~C.} \bibnamefont{Ahn}}, \bibinfo{journal}{npj
  2D Materials and Applications} \textbf{\bibinfo{volume}{4}},
  \bibinfo{pages}{17} (\bibinfo{year}{2020}).

\bibitem[{\citenamefont{Mi et~al.}(2023)\citenamefont{Mi, Xiao, Yu, Zhang,
  Wang, Cao, and Wang}}]{mi2023two}
\bibinfo{author}{\bibfnamefont{M.}~\bibnamefont{Mi}},
  \bibinfo{author}{\bibfnamefont{H.}~\bibnamefont{Xiao}},
  \bibinfo{author}{\bibfnamefont{L.}~\bibnamefont{Yu}},
  \bibinfo{author}{\bibfnamefont{Y.}~\bibnamefont{Zhang}},
  \bibinfo{author}{\bibfnamefont{Y.}~\bibnamefont{Wang}},
  \bibinfo{author}{\bibfnamefont{Q.}~\bibnamefont{Cao}}, \bibnamefont{and}
  \bibinfo{author}{\bibfnamefont{Y.}~\bibnamefont{Wang}},
  \bibinfo{journal}{Materials Today Nano} p. \bibinfo{pages}{100408}
  (\bibinfo{year}{2023}).

\bibitem[{\citenamefont{Liu et~al.}(2020)\citenamefont{Liu, Wang, Lu, Xie,
  Chen, Yin, Cheng, and Wu}}]{liu2020environmental}
\bibinfo{author}{\bibfnamefont{Y.}~\bibnamefont{Liu}},
  \bibinfo{author}{\bibfnamefont{W.}~\bibnamefont{Wang}},
  \bibinfo{author}{\bibfnamefont{H.}~\bibnamefont{Lu}},
  \bibinfo{author}{\bibfnamefont{Q.}~\bibnamefont{Xie}},
  \bibinfo{author}{\bibfnamefont{L.}~\bibnamefont{Chen}},
  \bibinfo{author}{\bibfnamefont{H.}~\bibnamefont{Yin}},
  \bibinfo{author}{\bibfnamefont{G.}~\bibnamefont{Cheng}}, \bibnamefont{and}
  \bibinfo{author}{\bibfnamefont{X.}~\bibnamefont{Wu}},
  \bibinfo{journal}{Applied Surface Science} \textbf{\bibinfo{volume}{511}},
  \bibinfo{pages}{145452} (\bibinfo{year}{2020}).

\bibitem[{\citenamefont{Galbiati et~al.}(2020)\citenamefont{Galbiati, Zatko,
  Godel, Hirschauer, Vecchiola, Bouzehouane, Collin, Servet, Cantarero, Petroff
  et~al.}}]{galbiati2020very}
\bibinfo{author}{\bibfnamefont{M.}~\bibnamefont{Galbiati}},
  \bibinfo{author}{\bibfnamefont{V.}~\bibnamefont{Zatko}},
  \bibinfo{author}{\bibfnamefont{F.}~\bibnamefont{Godel}},
  \bibinfo{author}{\bibfnamefont{P.}~\bibnamefont{Hirschauer}},
  \bibinfo{author}{\bibfnamefont{A.}~\bibnamefont{Vecchiola}},
  \bibinfo{author}{\bibfnamefont{K.}~\bibnamefont{Bouzehouane}},
  \bibinfo{author}{\bibfnamefont{S.}~\bibnamefont{Collin}},
  \bibinfo{author}{\bibfnamefont{B.}~\bibnamefont{Servet}},
  \bibinfo{author}{\bibfnamefont{A.}~\bibnamefont{Cantarero}},
  \bibinfo{author}{\bibfnamefont{F.}~\bibnamefont{Petroff}},
  \bibnamefont{et~al.}, \bibinfo{journal}{ACS Applied Electronic Materials}
  \textbf{\bibinfo{volume}{2}}, \bibinfo{pages}{3508} (\bibinfo{year}{2020}).

\bibitem[{\citenamefont{Gish et~al.}(2021)\citenamefont{Gish, Lebedev, Stanev,
  Jiang, Georgopoulos, Song, Lim, Garvey, Valdman, Balogun
  et~al.}}]{gish2021ambient}
\bibinfo{author}{\bibfnamefont{J.~T.} \bibnamefont{Gish}},
  \bibinfo{author}{\bibfnamefont{D.}~\bibnamefont{Lebedev}},
  \bibinfo{author}{\bibfnamefont{T.~K.} \bibnamefont{Stanev}},
  \bibinfo{author}{\bibfnamefont{S.}~\bibnamefont{Jiang}},
  \bibinfo{author}{\bibfnamefont{L.}~\bibnamefont{Georgopoulos}},
  \bibinfo{author}{\bibfnamefont{T.~W.} \bibnamefont{Song}},
  \bibinfo{author}{\bibfnamefont{G.}~\bibnamefont{Lim}},
  \bibinfo{author}{\bibfnamefont{E.~S.} \bibnamefont{Garvey}},
  \bibinfo{author}{\bibfnamefont{L.}~\bibnamefont{Valdman}},
  \bibinfo{author}{\bibfnamefont{O.}~\bibnamefont{Balogun}},
  \bibnamefont{et~al.}, \bibinfo{journal}{ACS nano}
  \textbf{\bibinfo{volume}{15}}, \bibinfo{pages}{10659} (\bibinfo{year}{2021}).

\bibitem[{\citenamefont{Shcherbakov et~al.}(2018)\citenamefont{Shcherbakov,
  Stepanov, Weber, Wang, Hu, Zhu, Watanabe, Taniguchi, Mao, Windl
  et~al.}}]{shcherbakov2018raman}
\bibinfo{author}{\bibfnamefont{D.}~\bibnamefont{Shcherbakov}},
  \bibinfo{author}{\bibfnamefont{P.}~\bibnamefont{Stepanov}},
  \bibinfo{author}{\bibfnamefont{D.}~\bibnamefont{Weber}},
  \bibinfo{author}{\bibfnamefont{Y.}~\bibnamefont{Wang}},
  \bibinfo{author}{\bibfnamefont{J.}~\bibnamefont{Hu}},
  \bibinfo{author}{\bibfnamefont{Y.}~\bibnamefont{Zhu}},
  \bibinfo{author}{\bibfnamefont{K.}~\bibnamefont{Watanabe}},
  \bibinfo{author}{\bibfnamefont{T.}~\bibnamefont{Taniguchi}},
  \bibinfo{author}{\bibfnamefont{Z.}~\bibnamefont{Mao}},
  \bibinfo{author}{\bibfnamefont{W.}~\bibnamefont{Windl}},
  \bibnamefont{et~al.}, \bibinfo{journal}{Nano letters}
  \textbf{\bibinfo{volume}{18}}, \bibinfo{pages}{4214} (\bibinfo{year}{2018}).

\bibitem[{\citenamefont{Ziebel et~al.}(2024)\citenamefont{Ziebel, Feuer, Cox,
  Zhu, Dean, and Roy}}]{ziebel2024crsbr}
\bibinfo{author}{\bibfnamefont{M.~E.} \bibnamefont{Ziebel}},
  \bibinfo{author}{\bibfnamefont{M.~L.} \bibnamefont{Feuer}},
  \bibinfo{author}{\bibfnamefont{J.}~\bibnamefont{Cox}},
  \bibinfo{author}{\bibfnamefont{X.}~\bibnamefont{Zhu}},
  \bibinfo{author}{\bibfnamefont{C.~R.} \bibnamefont{Dean}}, \bibnamefont{and}
  \bibinfo{author}{\bibfnamefont{X.}~\bibnamefont{Roy}}, \bibinfo{journal}{Nano
  Letters} \textbf{\bibinfo{volume}{24}}, \bibinfo{pages}{4319}
  (\bibinfo{year}{2024}).

\bibitem[{\citenamefont{Ye et~al.}(2022)\citenamefont{Ye, Wang, Wu, Liu, Zhou,
  Wang, Soll, Sofer, Yue, Liu et~al.}}]{ye2022layer}
\bibinfo{author}{\bibfnamefont{C.}~\bibnamefont{Ye}},
  \bibinfo{author}{\bibfnamefont{C.}~\bibnamefont{Wang}},
  \bibinfo{author}{\bibfnamefont{Q.}~\bibnamefont{Wu}},
  \bibinfo{author}{\bibfnamefont{S.}~\bibnamefont{Liu}},
  \bibinfo{author}{\bibfnamefont{J.}~\bibnamefont{Zhou}},
  \bibinfo{author}{\bibfnamefont{G.}~\bibnamefont{Wang}},
  \bibinfo{author}{\bibfnamefont{A.}~\bibnamefont{Soll}},
  \bibinfo{author}{\bibfnamefont{Z.}~\bibnamefont{Sofer}},
  \bibinfo{author}{\bibfnamefont{M.}~\bibnamefont{Yue}},
  \bibinfo{author}{\bibfnamefont{X.}~\bibnamefont{Liu}}, \bibnamefont{et~al.},
  \bibinfo{journal}{ACS nano} \textbf{\bibinfo{volume}{16}},
  \bibinfo{pages}{11876} (\bibinfo{year}{2022}).

\bibitem[{\citenamefont{Bae et~al.}(2022)\citenamefont{Bae, Wang, Scheie, Xu,
  Chica, Diederich, Cenker, Ziebel, Bai, Ren et~al.}}]{bae2022exciton}
\bibinfo{author}{\bibfnamefont{Y.~J.} \bibnamefont{Bae}},
  \bibinfo{author}{\bibfnamefont{J.}~\bibnamefont{Wang}},
  \bibinfo{author}{\bibfnamefont{A.}~\bibnamefont{Scheie}},
  \bibinfo{author}{\bibfnamefont{J.}~\bibnamefont{Xu}},
  \bibinfo{author}{\bibfnamefont{D.~G.} \bibnamefont{Chica}},
  \bibinfo{author}{\bibfnamefont{G.~M.} \bibnamefont{Diederich}},
  \bibinfo{author}{\bibfnamefont{J.}~\bibnamefont{Cenker}},
  \bibinfo{author}{\bibfnamefont{M.~E.} \bibnamefont{Ziebel}},
  \bibinfo{author}{\bibfnamefont{Y.}~\bibnamefont{Bai}},
  \bibinfo{author}{\bibfnamefont{H.}~\bibnamefont{Ren}}, \bibnamefont{et~al.},
  \bibinfo{journal}{Nature} \textbf{\bibinfo{volume}{609}},
  \bibinfo{pages}{282} (\bibinfo{year}{2022}).

\bibitem[{\citenamefont{Diederich et~al.}(2023)\citenamefont{Diederich, Cenker,
  Ren, Fonseca, Chica, Bae, Zhu, Roy, Cao, Xiao et~al.}}]{diederich2023tunable}
\bibinfo{author}{\bibfnamefont{G.~M.} \bibnamefont{Diederich}},
  \bibinfo{author}{\bibfnamefont{J.}~\bibnamefont{Cenker}},
  \bibinfo{author}{\bibfnamefont{Y.}~\bibnamefont{Ren}},
  \bibinfo{author}{\bibfnamefont{J.}~\bibnamefont{Fonseca}},
  \bibinfo{author}{\bibfnamefont{D.~G.} \bibnamefont{Chica}},
  \bibinfo{author}{\bibfnamefont{Y.~J.} \bibnamefont{Bae}},
  \bibinfo{author}{\bibfnamefont{X.}~\bibnamefont{Zhu}},
  \bibinfo{author}{\bibfnamefont{X.}~\bibnamefont{Roy}},
  \bibinfo{author}{\bibfnamefont{T.}~\bibnamefont{Cao}},
  \bibinfo{author}{\bibfnamefont{D.}~\bibnamefont{Xiao}}, \bibnamefont{et~al.},
  \bibinfo{journal}{Nature Nanotechnology} \textbf{\bibinfo{volume}{18}},
  \bibinfo{pages}{23} (\bibinfo{year}{2023}).

\bibitem[{\citenamefont{Diederich et~al.}(2024)\citenamefont{Diederich, Nguyen,
  Cenker, Fonseca, Pumulo, Bae, Chica, Roy, Zhu, Xiao
  et~al.}}]{diederich2024exciton}
\bibinfo{author}{\bibfnamefont{G.~M.} \bibnamefont{Diederich}},
  \bibinfo{author}{\bibfnamefont{M.}~\bibnamefont{Nguyen}},
  \bibinfo{author}{\bibfnamefont{J.}~\bibnamefont{Cenker}},
  \bibinfo{author}{\bibfnamefont{J.}~\bibnamefont{Fonseca}},
  \bibinfo{author}{\bibfnamefont{S.}~\bibnamefont{Pumulo}},
  \bibinfo{author}{\bibfnamefont{Y.~J.} \bibnamefont{Bae}},
  \bibinfo{author}{\bibfnamefont{D.~G.} \bibnamefont{Chica}},
  \bibinfo{author}{\bibfnamefont{X.}~\bibnamefont{Roy}},
  \bibinfo{author}{\bibfnamefont{X.}~\bibnamefont{Zhu}},
  \bibinfo{author}{\bibfnamefont{D.}~\bibnamefont{Xiao}}, \bibnamefont{et~al.},
  \bibinfo{journal}{arXiv preprint arXiv:2411.14943}  (\bibinfo{year}{2024}).

\bibitem[{\citenamefont{Lin et~al.}(2024{\natexlab{a}})\citenamefont{Lin, Sun,
  Dirnberger, Li, Qu, Wen, Sofer, S\"oll, Winnerl, Helm
  et~al.}}]{lin2024strong}
\bibinfo{author}{\bibfnamefont{K.}~\bibnamefont{Lin}},
  \bibinfo{author}{\bibfnamefont{X.}~\bibnamefont{Sun}},
  \bibinfo{author}{\bibfnamefont{F.}~\bibnamefont{Dirnberger}},
  \bibinfo{author}{\bibfnamefont{Y.}~\bibnamefont{Li}},
  \bibinfo{author}{\bibfnamefont{J.}~\bibnamefont{Qu}},
  \bibinfo{author}{\bibfnamefont{P.}~\bibnamefont{Wen}},
  \bibinfo{author}{\bibfnamefont{Z.}~\bibnamefont{Sofer}},
  \bibinfo{author}{\bibfnamefont{A.}~\bibnamefont{S\"oll}},
  \bibinfo{author}{\bibfnamefont{S.}~\bibnamefont{Winnerl}},
  \bibinfo{author}{\bibfnamefont{M.}~\bibnamefont{Helm}}, \bibnamefont{et~al.},
  \bibinfo{journal}{ACS nano} \textbf{\bibinfo{volume}{18}},
  \bibinfo{pages}{2898} (\bibinfo{year}{2024}{\natexlab{a}}).

\bibitem[{\citenamefont{Mondal et~al.}(2024)\citenamefont{Mondal, Markina,
  Hopf, Krelle, Shradha, Klein, Glazov, Gerber, Hagmann, Klitzing
  et~al.}}]{mondal2024raman}
\bibinfo{author}{\bibfnamefont{P.}~\bibnamefont{Mondal}},
  \bibinfo{author}{\bibfnamefont{D.~I.} \bibnamefont{Markina}},
  \bibinfo{author}{\bibfnamefont{L.}~\bibnamefont{Hopf}},
  \bibinfo{author}{\bibfnamefont{L.}~\bibnamefont{Krelle}},
  \bibinfo{author}{\bibfnamefont{S.}~\bibnamefont{Shradha}},
  \bibinfo{author}{\bibfnamefont{J.}~\bibnamefont{Klein}},
  \bibinfo{author}{\bibfnamefont{M.~M.} \bibnamefont{Glazov}},
  \bibinfo{author}{\bibfnamefont{I.}~\bibnamefont{Gerber}},
  \bibinfo{author}{\bibfnamefont{K.}~\bibnamefont{Hagmann}},
  \bibinfo{author}{\bibfnamefont{R.~v.} \bibnamefont{Klitzing}},
  \bibnamefont{et~al.}, \bibinfo{journal}{arXiv preprint arXiv:2410.22164}
  (\bibinfo{year}{2024}).

\bibitem[{\citenamefont{Sahu et~al.}(2025)\citenamefont{Sahu,
  Berrezueta-Palacios, Juergensen, Mosina, Sofer, Velický, Kusch, and
  Frank}}]{sahu2025resonanceramanscatteringanomalous}
\bibinfo{author}{\bibfnamefont{S.}~\bibnamefont{Sahu}},
  \bibinfo{author}{\bibfnamefont{C.}~\bibnamefont{Berrezueta-Palacios}},
  \bibinfo{author}{\bibfnamefont{S.}~\bibnamefont{Juergensen}},
  \bibinfo{author}{\bibfnamefont{K.}~\bibnamefont{Mosina}},
  \bibinfo{author}{\bibfnamefont{Z.}~\bibnamefont{Sofer}},
  \bibinfo{author}{\bibfnamefont{M.}~\bibnamefont{Velický}},
  \bibinfo{author}{\bibfnamefont{P.}~\bibnamefont{Kusch}}, \bibnamefont{and}
  \bibinfo{author}{\bibfnamefont{O.}~\bibnamefont{Frank}},
  \bibinfo{journal}{arXiv preprint arXiv:2502.01794}  (\bibinfo{year}{2025}).

\bibitem[{\citenamefont{Dirnberger et~al.}(2023)\citenamefont{Dirnberger, Quan,
  Bushati, Diederich, Florian, Klein, Mosina, Sofer, Xu, Kamra
  et~al.}}]{dirnberger2023magneto}
\bibinfo{author}{\bibfnamefont{F.}~\bibnamefont{Dirnberger}},
  \bibinfo{author}{\bibfnamefont{J.}~\bibnamefont{Quan}},
  \bibinfo{author}{\bibfnamefont{R.}~\bibnamefont{Bushati}},
  \bibinfo{author}{\bibfnamefont{G.~M.} \bibnamefont{Diederich}},
  \bibinfo{author}{\bibfnamefont{M.}~\bibnamefont{Florian}},
  \bibinfo{author}{\bibfnamefont{J.}~\bibnamefont{Klein}},
  \bibinfo{author}{\bibfnamefont{K.}~\bibnamefont{Mosina}},
  \bibinfo{author}{\bibfnamefont{Z.}~\bibnamefont{Sofer}},
  \bibinfo{author}{\bibfnamefont{X.}~\bibnamefont{Xu}},
  \bibinfo{author}{\bibfnamefont{A.}~\bibnamefont{Kamra}},
  \bibnamefont{et~al.}, \bibinfo{journal}{Nature}
  \textbf{\bibinfo{volume}{620}}, \bibinfo{pages}{533} (\bibinfo{year}{2023}).

\bibitem[{\citenamefont{Wang et~al.}(2023)\citenamefont{Wang, Zhang, Yang, Lin,
  Chen, Yang, Gong, Chen, Ye, and Liu}}]{wang2023magnetically}
\bibinfo{author}{\bibfnamefont{T.}~\bibnamefont{Wang}},
  \bibinfo{author}{\bibfnamefont{D.}~\bibnamefont{Zhang}},
  \bibinfo{author}{\bibfnamefont{S.}~\bibnamefont{Yang}},
  \bibinfo{author}{\bibfnamefont{Z.}~\bibnamefont{Lin}},
  \bibinfo{author}{\bibfnamefont{Q.}~\bibnamefont{Chen}},
  \bibinfo{author}{\bibfnamefont{J.}~\bibnamefont{Yang}},
  \bibinfo{author}{\bibfnamefont{Q.}~\bibnamefont{Gong}},
  \bibinfo{author}{\bibfnamefont{Z.}~\bibnamefont{Chen}},
  \bibinfo{author}{\bibfnamefont{Y.}~\bibnamefont{Ye}}, \bibnamefont{and}
  \bibinfo{author}{\bibfnamefont{W.}~\bibnamefont{Liu}},
  \bibinfo{journal}{Nature Communications} \textbf{\bibinfo{volume}{14}},
  \bibinfo{pages}{5966} (\bibinfo{year}{2023}).

\bibitem[{\citenamefont{Li et~al.}(2024)\citenamefont{Li, Shen, Jiang, Tang,
  Liu, Guo, Liang, Song, Deng, and Zhang}}]{li20242d}
\bibinfo{author}{\bibfnamefont{C.}~\bibnamefont{Li}},
  \bibinfo{author}{\bibfnamefont{C.}~\bibnamefont{Shen}},
  \bibinfo{author}{\bibfnamefont{N.}~\bibnamefont{Jiang}},
  \bibinfo{author}{\bibfnamefont{K.~K.} \bibnamefont{Tang}},
  \bibinfo{author}{\bibfnamefont{X.}~\bibnamefont{Liu}},
  \bibinfo{author}{\bibfnamefont{J.}~\bibnamefont{Guo}},
  \bibinfo{author}{\bibfnamefont{Y.}~\bibnamefont{Liang}},
  \bibinfo{author}{\bibfnamefont{J.}~\bibnamefont{Song}},
  \bibinfo{author}{\bibfnamefont{X.}~\bibnamefont{Deng}}, \bibnamefont{and}
  \bibinfo{author}{\bibfnamefont{Q.}~\bibnamefont{Zhang}},
  \bibinfo{journal}{Advanced Functional Materials} p. \bibinfo{pages}{2411589}
  (\bibinfo{year}{2024}).

\bibitem[{\citenamefont{Nessi et~al.}(2024)\citenamefont{Nessi, Occhialini,
  Demir, Powalla, and Comin}}]{nessi2024magnetic}
\bibinfo{author}{\bibfnamefont{L.}~\bibnamefont{Nessi}},
  \bibinfo{author}{\bibfnamefont{C.~A.} \bibnamefont{Occhialini}},
  \bibinfo{author}{\bibfnamefont{A.~K.} \bibnamefont{Demir}},
  \bibinfo{author}{\bibfnamefont{L.}~\bibnamefont{Powalla}}, \bibnamefont{and}
  \bibinfo{author}{\bibfnamefont{R.}~\bibnamefont{Comin}},
  \bibinfo{journal}{ACS nano}  (\bibinfo{year}{2024}).

\bibitem[{\citenamefont{Han et~al.}(2025)\citenamefont{Han, Shan, Song,
  Lackner, Esmann, Solovyeva, Eilenberger, Regner, Sofer, Kyriienko
  et~al.}}]{han2025exciton}
\bibinfo{author}{\bibfnamefont{B.}~\bibnamefont{Han}},
  \bibinfo{author}{\bibfnamefont{H.}~\bibnamefont{Shan}},
  \bibinfo{author}{\bibfnamefont{K.~W.} \bibnamefont{Song}},
  \bibinfo{author}{\bibfnamefont{L.}~\bibnamefont{Lackner}},
  \bibinfo{author}{\bibfnamefont{M.}~\bibnamefont{Esmann}},
  \bibinfo{author}{\bibfnamefont{V.}~\bibnamefont{Solovyeva}},
  \bibinfo{author}{\bibfnamefont{F.}~\bibnamefont{Eilenberger}},
  \bibinfo{author}{\bibfnamefont{J.}~\bibnamefont{Regner}},
  \bibinfo{author}{\bibfnamefont{Z.}~\bibnamefont{Sofer}},
  \bibinfo{author}{\bibfnamefont{O.}~\bibnamefont{Kyriienko}},
  \bibnamefont{et~al.}, \bibinfo{journal}{arXiv preprint arXiv:2501.18233}
  (\bibinfo{year}{2025}).

\bibitem[{\citenamefont{Boix-Constant et~al.}(2022)\citenamefont{Boix-Constant,
  Ma{\~n}as-Valero, Ruiz, Rybakov, Konieczny, Pillet, Baldov{\'\i}, and
  Coronado}}]{boix2022probing}
\bibinfo{author}{\bibfnamefont{C.}~\bibnamefont{Boix-Constant}},
  \bibinfo{author}{\bibfnamefont{S.}~\bibnamefont{Ma{\~n}as-Valero}},
  \bibinfo{author}{\bibfnamefont{A.~M.} \bibnamefont{Ruiz}},
  \bibinfo{author}{\bibfnamefont{A.}~\bibnamefont{Rybakov}},
  \bibinfo{author}{\bibfnamefont{K.~A.} \bibnamefont{Konieczny}},
  \bibinfo{author}{\bibfnamefont{S.}~\bibnamefont{Pillet}},
  \bibinfo{author}{\bibfnamefont{J.~J.} \bibnamefont{Baldov{\'\i}}},
  \bibnamefont{and} \bibinfo{author}{\bibfnamefont{E.}~\bibnamefont{Coronado}},
  \bibinfo{journal}{Advanced Materials} \textbf{\bibinfo{volume}{34}},
  \bibinfo{pages}{2204940} (\bibinfo{year}{2022}).

\bibitem[{\citenamefont{Telford et~al.}(2020)\citenamefont{Telford, Dismukes,
  Lee, Cheng, Wieteska, Bartholomew, Chen, Xu, Pasupathy, Zhu
  et~al.}}]{telford2020layered}
\bibinfo{author}{\bibfnamefont{E.~J.} \bibnamefont{Telford}},
  \bibinfo{author}{\bibfnamefont{A.~H.} \bibnamefont{Dismukes}},
  \bibinfo{author}{\bibfnamefont{K.}~\bibnamefont{Lee}},
  \bibinfo{author}{\bibfnamefont{M.}~\bibnamefont{Cheng}},
  \bibinfo{author}{\bibfnamefont{A.}~\bibnamefont{Wieteska}},
  \bibinfo{author}{\bibfnamefont{A.~K.} \bibnamefont{Bartholomew}},
  \bibinfo{author}{\bibfnamefont{Y.-S.} \bibnamefont{Chen}},
  \bibinfo{author}{\bibfnamefont{X.}~\bibnamefont{Xu}},
  \bibinfo{author}{\bibfnamefont{A.~N.} \bibnamefont{Pasupathy}},
  \bibinfo{author}{\bibfnamefont{X.}~\bibnamefont{Zhu}}, \bibnamefont{et~al.},
  \bibinfo{journal}{Advanced Materials} \textbf{\bibinfo{volume}{32}},
  \bibinfo{pages}{2003240} (\bibinfo{year}{2020}).

\bibitem[{\citenamefont{Wilson et~al.}(2021)\citenamefont{Wilson, Lee, Cenker,
  Xie, Dismukes, Telford, Fonseca, Sivakumar, Dean, Cao
  et~al.}}]{wilson2021interlayer}
\bibinfo{author}{\bibfnamefont{N.~P.} \bibnamefont{Wilson}},
  \bibinfo{author}{\bibfnamefont{K.}~\bibnamefont{Lee}},
  \bibinfo{author}{\bibfnamefont{J.}~\bibnamefont{Cenker}},
  \bibinfo{author}{\bibfnamefont{K.}~\bibnamefont{Xie}},
  \bibinfo{author}{\bibfnamefont{A.~H.} \bibnamefont{Dismukes}},
  \bibinfo{author}{\bibfnamefont{E.~J.} \bibnamefont{Telford}},
  \bibinfo{author}{\bibfnamefont{J.}~\bibnamefont{Fonseca}},
  \bibinfo{author}{\bibfnamefont{S.}~\bibnamefont{Sivakumar}},
  \bibinfo{author}{\bibfnamefont{C.}~\bibnamefont{Dean}},
  \bibinfo{author}{\bibfnamefont{T.}~\bibnamefont{Cao}}, \bibnamefont{et~al.},
  \bibinfo{journal}{Nature Materials} \textbf{\bibinfo{volume}{20}},
  \bibinfo{pages}{1657} (\bibinfo{year}{2021}).

\bibitem[{\citenamefont{Tabataba-Vakili
  et~al.}(2024)\citenamefont{Tabataba-Vakili, Nguyen, Rupp, Mosina,
  Papavasileiou, Watanabe, Taniguchi, Maletinsky, Glazov, Sofer
  et~al.}}]{tabataba2024doping}
\bibinfo{author}{\bibfnamefont{F.}~\bibnamefont{Tabataba-Vakili}},
  \bibinfo{author}{\bibfnamefont{H.~P.} \bibnamefont{Nguyen}},
  \bibinfo{author}{\bibfnamefont{A.}~\bibnamefont{Rupp}},
  \bibinfo{author}{\bibfnamefont{K.}~\bibnamefont{Mosina}},
  \bibinfo{author}{\bibfnamefont{A.}~\bibnamefont{Papavasileiou}},
  \bibinfo{author}{\bibfnamefont{K.}~\bibnamefont{Watanabe}},
  \bibinfo{author}{\bibfnamefont{T.}~\bibnamefont{Taniguchi}},
  \bibinfo{author}{\bibfnamefont{P.}~\bibnamefont{Maletinsky}},
  \bibinfo{author}{\bibfnamefont{M.~M.} \bibnamefont{Glazov}},
  \bibinfo{author}{\bibfnamefont{Z.}~\bibnamefont{Sofer}},
  \bibnamefont{et~al.}, \bibinfo{journal}{Nature Communications}
  \textbf{\bibinfo{volume}{15}}, \bibinfo{pages}{4735} (\bibinfo{year}{2024}).

\bibitem[{\citenamefont{Klein et~al.}(2023)\citenamefont{Klein, Pingault,
  Florian, Hei{\ss}enb\"uttel, Steinhoff, Song, Torres, Dirnberger, Curtis,
  Weile et~al.}}]{klein2023bulk}
\bibinfo{author}{\bibfnamefont{J.}~\bibnamefont{Klein}},
  \bibinfo{author}{\bibfnamefont{B.}~\bibnamefont{Pingault}},
  \bibinfo{author}{\bibfnamefont{M.}~\bibnamefont{Florian}},
  \bibinfo{author}{\bibfnamefont{M.-C.} \bibnamefont{Hei{\ss}enb\"uttel}},
  \bibinfo{author}{\bibfnamefont{A.}~\bibnamefont{Steinhoff}},
  \bibinfo{author}{\bibfnamefont{Z.}~\bibnamefont{Song}},
  \bibinfo{author}{\bibfnamefont{K.}~\bibnamefont{Torres}},
  \bibinfo{author}{\bibfnamefont{F.}~\bibnamefont{Dirnberger}},
  \bibinfo{author}{\bibfnamefont{J.~B.} \bibnamefont{Curtis}},
  \bibinfo{author}{\bibfnamefont{M.}~\bibnamefont{Weile}},
  \bibnamefont{et~al.}, \bibinfo{journal}{ACS nano}
  \textbf{\bibinfo{volume}{17}}, \bibinfo{pages}{5316} (\bibinfo{year}{2023}).

\bibitem[{\citenamefont{Yang et~al.}(2021)\citenamefont{Yang, Wang, Liu, Lu,
  and Wu}}]{yang2021triaxial}
\bibinfo{author}{\bibfnamefont{K.}~\bibnamefont{Yang}},
  \bibinfo{author}{\bibfnamefont{G.}~\bibnamefont{Wang}},
  \bibinfo{author}{\bibfnamefont{L.}~\bibnamefont{Liu}},
  \bibinfo{author}{\bibfnamefont{D.}~\bibnamefont{Lu}}, \bibnamefont{and}
  \bibinfo{author}{\bibfnamefont{H.}~\bibnamefont{Wu}},
  \bibinfo{journal}{Physical Review B} \textbf{\bibinfo{volume}{104}},
  \bibinfo{pages}{144416} (\bibinfo{year}{2021}).

\bibitem[{\citenamefont{Liebich et~al.}(2025)\citenamefont{Liebich, Florian,
  Nilforoushan, Mooshammer, Koulouklidis, Wittmann, Mosina, Sofer, Dirnberger,
  Kira et~al.}}]{liebich2025controlling}
\bibinfo{author}{\bibfnamefont{M.}~\bibnamefont{Liebich}},
  \bibinfo{author}{\bibfnamefont{M.}~\bibnamefont{Florian}},
  \bibinfo{author}{\bibfnamefont{N.}~\bibnamefont{Nilforoushan}},
  \bibinfo{author}{\bibfnamefont{F.}~\bibnamefont{Mooshammer}},
  \bibinfo{author}{\bibfnamefont{A.}~\bibnamefont{Koulouklidis}},
  \bibinfo{author}{\bibfnamefont{L.}~\bibnamefont{Wittmann}},
  \bibinfo{author}{\bibfnamefont{K.}~\bibnamefont{Mosina}},
  \bibinfo{author}{\bibfnamefont{Z.}~\bibnamefont{Sofer}},
  \bibinfo{author}{\bibfnamefont{F.}~\bibnamefont{Dirnberger}},
  \bibinfo{author}{\bibfnamefont{M.}~\bibnamefont{Kira}}, \bibnamefont{et~al.},
  \bibinfo{journal}{Nature Materials}  (\bibinfo{year}{2025}).

\bibitem[{\citenamefont{Shao et~al.}(2025)\citenamefont{Shao, Dirnberger, Qiu,
  Acharya, Terres, Telford, Pashov, Kim, Ruta, Chica
  et~al.}}]{shao2025magnetically}
\bibinfo{author}{\bibfnamefont{Y.}~\bibnamefont{Shao}},
  \bibinfo{author}{\bibfnamefont{F.}~\bibnamefont{Dirnberger}},
  \bibinfo{author}{\bibfnamefont{S.}~\bibnamefont{Qiu}},
  \bibinfo{author}{\bibfnamefont{S.}~\bibnamefont{Acharya}},
  \bibinfo{author}{\bibfnamefont{S.}~\bibnamefont{Terres}},
  \bibinfo{author}{\bibfnamefont{E.~J.} \bibnamefont{Telford}},
  \bibinfo{author}{\bibfnamefont{D.}~\bibnamefont{Pashov}},
  \bibinfo{author}{\bibfnamefont{B.~S.} \bibnamefont{Kim}},
  \bibinfo{author}{\bibfnamefont{F.~L.} \bibnamefont{Ruta}},
  \bibinfo{author}{\bibfnamefont{D.~G.} \bibnamefont{Chica}},
  \bibnamefont{et~al.}, \bibinfo{journal}{Nature materials}
  (\bibinfo{year}{2025}).

\bibitem[{\citenamefont{Tschudin et~al.}(2024)\citenamefont{Tschudin, Broadway,
  Siegwolf, Schrader, Telford, Gross, Cox, Dubois, Chica, Rama-Eiroa
  et~al.}}]{tschudin2024imaging}
\bibinfo{author}{\bibfnamefont{M.~A.} \bibnamefont{Tschudin}},
  \bibinfo{author}{\bibfnamefont{D.~A.} \bibnamefont{Broadway}},
  \bibinfo{author}{\bibfnamefont{P.}~\bibnamefont{Siegwolf}},
  \bibinfo{author}{\bibfnamefont{C.}~\bibnamefont{Schrader}},
  \bibinfo{author}{\bibfnamefont{E.~J.} \bibnamefont{Telford}},
  \bibinfo{author}{\bibfnamefont{B.}~\bibnamefont{Gross}},
  \bibinfo{author}{\bibfnamefont{J.}~\bibnamefont{Cox}},
  \bibinfo{author}{\bibfnamefont{A.~E.} \bibnamefont{Dubois}},
  \bibinfo{author}{\bibfnamefont{D.~G.} \bibnamefont{Chica}},
  \bibinfo{author}{\bibfnamefont{R.}~\bibnamefont{Rama-Eiroa}},
  \bibnamefont{et~al.}, \bibinfo{journal}{Nature Communications}
  \textbf{\bibinfo{volume}{15}}, \bibinfo{pages}{6005} (\bibinfo{year}{2024}).

\bibitem[{\citenamefont{Rizzo et~al.}(2022)\citenamefont{Rizzo, McLeod,
  Carnahan, Telford, Dismukes, Wiscons, Dong, Nuckolls, Dean, Pasupathy
  et~al.}}]{rizzo2022visualizing}
\bibinfo{author}{\bibfnamefont{D.~J.} \bibnamefont{Rizzo}},
  \bibinfo{author}{\bibfnamefont{A.~S.} \bibnamefont{McLeod}},
  \bibinfo{author}{\bibfnamefont{C.}~\bibnamefont{Carnahan}},
  \bibinfo{author}{\bibfnamefont{E.~J.} \bibnamefont{Telford}},
  \bibinfo{author}{\bibfnamefont{A.~H.} \bibnamefont{Dismukes}},
  \bibinfo{author}{\bibfnamefont{R.~A.} \bibnamefont{Wiscons}},
  \bibinfo{author}{\bibfnamefont{Y.}~\bibnamefont{Dong}},
  \bibinfo{author}{\bibfnamefont{C.}~\bibnamefont{Nuckolls}},
  \bibinfo{author}{\bibfnamefont{C.~R.} \bibnamefont{Dean}},
  \bibinfo{author}{\bibfnamefont{A.~N.} \bibnamefont{Pasupathy}},
  \bibnamefont{et~al.}, \bibinfo{journal}{Advanced Materials}
  \textbf{\bibinfo{volume}{34}}, \bibinfo{pages}{2201000}
  (\bibinfo{year}{2022}).

\bibitem[{\citenamefont{Lee et~al.}(2021)\citenamefont{Lee, Dismukes, Telford,
  Wiscons, Wang, Xu, Nuckolls, Dean, Roy, and Zhu}}]{lee2021magnetic}
\bibinfo{author}{\bibfnamefont{K.}~\bibnamefont{Lee}},
  \bibinfo{author}{\bibfnamefont{A.~H.} \bibnamefont{Dismukes}},
  \bibinfo{author}{\bibfnamefont{E.~J.} \bibnamefont{Telford}},
  \bibinfo{author}{\bibfnamefont{R.~A.} \bibnamefont{Wiscons}},
  \bibinfo{author}{\bibfnamefont{J.}~\bibnamefont{Wang}},
  \bibinfo{author}{\bibfnamefont{X.}~\bibnamefont{Xu}},
  \bibinfo{author}{\bibfnamefont{C.}~\bibnamefont{Nuckolls}},
  \bibinfo{author}{\bibfnamefont{C.~R.} \bibnamefont{Dean}},
  \bibinfo{author}{\bibfnamefont{X.}~\bibnamefont{Roy}}, \bibnamefont{and}
  \bibinfo{author}{\bibfnamefont{X.}~\bibnamefont{Zhu}}, \bibinfo{journal}{Nano
  Letters} \textbf{\bibinfo{volume}{21}}, \bibinfo{pages}{3511}
  (\bibinfo{year}{2021}).

\bibitem[{\citenamefont{G{\"o}ser et~al.}(1990)\citenamefont{G{\"o}ser, Paul,
  and Kahle}}]{goser1990magnetic}
\bibinfo{author}{\bibfnamefont{O.}~\bibnamefont{G{\"o}ser}},
  \bibinfo{author}{\bibfnamefont{W.}~\bibnamefont{Paul}}, \bibnamefont{and}
  \bibinfo{author}{\bibfnamefont{H.}~\bibnamefont{Kahle}},
  \bibinfo{journal}{Journal of magnetism and magnetic materials}
  \textbf{\bibinfo{volume}{92}}, \bibinfo{pages}{129} (\bibinfo{year}{1990}).

\bibitem[{\citenamefont{Hei{\ss}enb\"uttel
  et~al.}(2025)\citenamefont{Hei{\ss}enb\"uttel, Piel, Klein, Deilmann,
  Wurstbauer, and Rohlfing}}]{PhysRevB.111.075107}
\bibinfo{author}{\bibfnamefont{M.-C.} \bibnamefont{Hei{\ss}enb\"uttel}},
  \bibinfo{author}{\bibfnamefont{P.-M.} \bibnamefont{Piel}},
  \bibinfo{author}{\bibfnamefont{J.}~\bibnamefont{Klein}},
  \bibinfo{author}{\bibfnamefont{T.}~\bibnamefont{Deilmann}},
  \bibinfo{author}{\bibfnamefont{U.}~\bibnamefont{Wurstbauer}},
  \bibnamefont{and} \bibinfo{author}{\bibfnamefont{M.}~\bibnamefont{Rohlfing}},
  \bibinfo{journal}{Phys. Rev. B} \textbf{\bibinfo{volume}{111}},
  \bibinfo{pages}{075107} (\bibinfo{year}{2025}).

\bibitem[{\citenamefont{Wang et~al.}(2017)\citenamefont{Wang, Robert, Glazov,
  Cadiz, Courtade, Amand, Lagarde, Taniguchi, Watanabe, Urbaszek
  et~al.}}]{wang2017plane}
\bibinfo{author}{\bibfnamefont{G.}~\bibnamefont{Wang}},
  \bibinfo{author}{\bibfnamefont{C.}~\bibnamefont{Robert}},
  \bibinfo{author}{\bibfnamefont{M.~M.} \bibnamefont{Glazov}},
  \bibinfo{author}{\bibfnamefont{F.}~\bibnamefont{Cadiz}},
  \bibinfo{author}{\bibfnamefont{E.}~\bibnamefont{Courtade}},
  \bibinfo{author}{\bibfnamefont{T.}~\bibnamefont{Amand}},
  \bibinfo{author}{\bibfnamefont{D.}~\bibnamefont{Lagarde}},
  \bibinfo{author}{\bibfnamefont{T.}~\bibnamefont{Taniguchi}},
  \bibinfo{author}{\bibfnamefont{K.}~\bibnamefont{Watanabe}},
  \bibinfo{author}{\bibfnamefont{B.}~\bibnamefont{Urbaszek}},
  \bibnamefont{et~al.}, \bibinfo{journal}{Physical review letters}
  \textbf{\bibinfo{volume}{119}}, \bibinfo{pages}{047401}
  (\bibinfo{year}{2017}).

\bibitem[{\citenamefont{He et~al.}(2020)\citenamefont{He, Rivera, Van~Tuan,
  Wilson, Yang, Taniguchi, Watanabe, Yan, Mandrus, Yu et~al.}}]{he2020valley}
\bibinfo{author}{\bibfnamefont{M.}~\bibnamefont{He}},
  \bibinfo{author}{\bibfnamefont{P.}~\bibnamefont{Rivera}},
  \bibinfo{author}{\bibfnamefont{D.}~\bibnamefont{Van~Tuan}},
  \bibinfo{author}{\bibfnamefont{N.~P.} \bibnamefont{Wilson}},
  \bibinfo{author}{\bibfnamefont{M.}~\bibnamefont{Yang}},
  \bibinfo{author}{\bibfnamefont{T.}~\bibnamefont{Taniguchi}},
  \bibinfo{author}{\bibfnamefont{K.}~\bibnamefont{Watanabe}},
  \bibinfo{author}{\bibfnamefont{J.}~\bibnamefont{Yan}},
  \bibinfo{author}{\bibfnamefont{D.~G.} \bibnamefont{Mandrus}},
  \bibinfo{author}{\bibfnamefont{H.}~\bibnamefont{Yu}}, \bibnamefont{et~al.},
  \bibinfo{journal}{Nature communications} \textbf{\bibinfo{volume}{11}},
  \bibinfo{pages}{618} (\bibinfo{year}{2020}).

\bibitem[{\citenamefont{Shree et~al.}(2021)\citenamefont{Shree, Paradisanos,
  Marie, Robert, and Urbaszek}}]{shree2021guide}
\bibinfo{author}{\bibfnamefont{S.}~\bibnamefont{Shree}},
  \bibinfo{author}{\bibfnamefont{I.}~\bibnamefont{Paradisanos}},
  \bibinfo{author}{\bibfnamefont{X.}~\bibnamefont{Marie}},
  \bibinfo{author}{\bibfnamefont{C.}~\bibnamefont{Robert}}, \bibnamefont{and}
  \bibinfo{author}{\bibfnamefont{B.}~\bibnamefont{Urbaszek}},
  \bibinfo{journal}{Nature Reviews Physics} \textbf{\bibinfo{volume}{3}},
  \bibinfo{pages}{39} (\bibinfo{year}{2021}).

\bibitem[{\citenamefont{Raja et~al.}(2017)\citenamefont{Raja, Chaves, Yu,
  Arefe, Hill, Rigosi, Berkelbach, Nagler, Sch{\"u}ller, Korn
  et~al.}}]{raja2017coulomb}
\bibinfo{author}{\bibfnamefont{A.}~\bibnamefont{Raja}},
  \bibinfo{author}{\bibfnamefont{A.}~\bibnamefont{Chaves}},
  \bibinfo{author}{\bibfnamefont{J.}~\bibnamefont{Yu}},
  \bibinfo{author}{\bibfnamefont{G.}~\bibnamefont{Arefe}},
  \bibinfo{author}{\bibfnamefont{H.~M.} \bibnamefont{Hill}},
  \bibinfo{author}{\bibfnamefont{A.~F.} \bibnamefont{Rigosi}},
  \bibinfo{author}{\bibfnamefont{T.~C.} \bibnamefont{Berkelbach}},
  \bibinfo{author}{\bibfnamefont{P.}~\bibnamefont{Nagler}},
  \bibinfo{author}{\bibfnamefont{C.}~\bibnamefont{Sch{\"u}ller}},
  \bibinfo{author}{\bibfnamefont{T.}~\bibnamefont{Korn}}, \bibnamefont{et~al.},
  \bibinfo{journal}{Nature communications} \textbf{\bibinfo{volume}{8}},
  \bibinfo{pages}{15251} (\bibinfo{year}{2017}).

\bibitem[{\citenamefont{Stier et~al.}(2016)\citenamefont{Stier, Wilson, Clark,
  Xu, and Crooker}}]{stier2016probing}
\bibinfo{author}{\bibfnamefont{A.~V.} \bibnamefont{Stier}},
  \bibinfo{author}{\bibfnamefont{N.~P.} \bibnamefont{Wilson}},
  \bibinfo{author}{\bibfnamefont{G.}~\bibnamefont{Clark}},
  \bibinfo{author}{\bibfnamefont{X.}~\bibnamefont{Xu}}, \bibnamefont{and}
  \bibinfo{author}{\bibfnamefont{S.~A.} \bibnamefont{Crooker}},
  \bibinfo{journal}{Nano letters} \textbf{\bibinfo{volume}{16}},
  \bibinfo{pages}{7054} (\bibinfo{year}{2016}).

\bibitem[{\citenamefont{Bianchi et~al.}(2023)\citenamefont{Bianchi, Acharya,
  Dirnberger, Klein, Pashov, Mosina, Sofer, Rudenko, Katsnelson,
  Van~Schilfgaarde et~al.}}]{bianchi2023paramagnetic}
\bibinfo{author}{\bibfnamefont{M.}~\bibnamefont{Bianchi}},
  \bibinfo{author}{\bibfnamefont{S.}~\bibnamefont{Acharya}},
  \bibinfo{author}{\bibfnamefont{F.}~\bibnamefont{Dirnberger}},
  \bibinfo{author}{\bibfnamefont{J.}~\bibnamefont{Klein}},
  \bibinfo{author}{\bibfnamefont{D.}~\bibnamefont{Pashov}},
  \bibinfo{author}{\bibfnamefont{K.}~\bibnamefont{Mosina}},
  \bibinfo{author}{\bibfnamefont{Z.}~\bibnamefont{Sofer}},
  \bibinfo{author}{\bibfnamefont{A.~N.} \bibnamefont{Rudenko}},
  \bibinfo{author}{\bibfnamefont{M.~I.} \bibnamefont{Katsnelson}},
  \bibinfo{author}{\bibfnamefont{M.}~\bibnamefont{Van~Schilfgaarde}},
  \bibnamefont{et~al.}, \bibinfo{journal}{Physical Review B}
  \textbf{\bibinfo{volume}{107}}, \bibinfo{pages}{235107}
  (\bibinfo{year}{2023}).

\bibitem[{\citenamefont{Lin et~al.}(2024{\natexlab{b}})\citenamefont{Lin, Li,
  Ghorbani-Asl, Sofer, Winnerl, Erbe, Krasheninnikov, Helm, Zhou, Dan
  et~al.}}]{lin2024probing}
\bibinfo{author}{\bibfnamefont{K.}~\bibnamefont{Lin}},
  \bibinfo{author}{\bibfnamefont{Y.}~\bibnamefont{Li}},
  \bibinfo{author}{\bibfnamefont{M.}~\bibnamefont{Ghorbani-Asl}},
  \bibinfo{author}{\bibfnamefont{Z.}~\bibnamefont{Sofer}},
  \bibinfo{author}{\bibfnamefont{S.}~\bibnamefont{Winnerl}},
  \bibinfo{author}{\bibfnamefont{A.}~\bibnamefont{Erbe}},
  \bibinfo{author}{\bibfnamefont{A.~V.} \bibnamefont{Krasheninnikov}},
  \bibinfo{author}{\bibfnamefont{M.}~\bibnamefont{Helm}},
  \bibinfo{author}{\bibfnamefont{S.}~\bibnamefont{Zhou}},
  \bibinfo{author}{\bibfnamefont{Y.}~\bibnamefont{Dan}}, \bibnamefont{et~al.},
  \bibinfo{journal}{The Journal of Physical Chemistry Letters}
  \textbf{\bibinfo{volume}{15}}, \bibinfo{pages}{6010}
  (\bibinfo{year}{2024}{\natexlab{b}}).

\bibitem[{\citenamefont{Komar et~al.}(2024)\citenamefont{Komar, {\L}opion,
  Goryca, Rybak, Wo{\'z}niak, Mosina, S{\"o}ll, Sofer, Pacuski, Faugeras
  et~al.}}]{komar2024colossal}
\bibinfo{author}{\bibfnamefont{R.}~\bibnamefont{Komar}},
  \bibinfo{author}{\bibfnamefont{A.}~\bibnamefont{{\L}opion}},
  \bibinfo{author}{\bibfnamefont{M.}~\bibnamefont{Goryca}},
  \bibinfo{author}{\bibfnamefont{M.}~\bibnamefont{Rybak}},
  \bibinfo{author}{\bibfnamefont{T.}~\bibnamefont{Wo{\'z}niak}},
  \bibinfo{author}{\bibfnamefont{K.}~\bibnamefont{Mosina}},
  \bibinfo{author}{\bibfnamefont{A.}~\bibnamefont{S{\"o}ll}},
  \bibinfo{author}{\bibfnamefont{Z.}~\bibnamefont{Sofer}},
  \bibinfo{author}{\bibfnamefont{W.}~\bibnamefont{Pacuski}},
  \bibinfo{author}{\bibfnamefont{C.}~\bibnamefont{Faugeras}},
  \bibnamefont{et~al.}, \bibinfo{journal}{arXiv preprint arXiv:2409.00187}
  (\bibinfo{year}{2024}).

\bibitem[{\citenamefont{Smolenski et~al.}(2025)\citenamefont{Smolenski, Wen,
  Li, Downey, Alfrey, Liu, Kondusamy, Bostwick, Jozwiak, Rotenberg
  et~al.}}]{smolenski2025large}
\bibinfo{author}{\bibfnamefont{S.}~\bibnamefont{Smolenski}},
  \bibinfo{author}{\bibfnamefont{M.}~\bibnamefont{Wen}},
  \bibinfo{author}{\bibfnamefont{Q.}~\bibnamefont{Li}},
  \bibinfo{author}{\bibfnamefont{E.}~\bibnamefont{Downey}},
  \bibinfo{author}{\bibfnamefont{A.}~\bibnamefont{Alfrey}},
  \bibinfo{author}{\bibfnamefont{W.}~\bibnamefont{Liu}},
  \bibinfo{author}{\bibfnamefont{A.~L.} \bibnamefont{Kondusamy}},
  \bibinfo{author}{\bibfnamefont{A.}~\bibnamefont{Bostwick}},
  \bibinfo{author}{\bibfnamefont{C.}~\bibnamefont{Jozwiak}},
  \bibinfo{author}{\bibfnamefont{E.}~\bibnamefont{Rotenberg}},
  \bibnamefont{et~al.}, \bibinfo{journal}{Nature Communications}
  \textbf{\bibinfo{volume}{16}}, \bibinfo{pages}{1134} (\bibinfo{year}{2025}).

\bibitem[{\citenamefont{Qian et~al.}(2023)\citenamefont{Qian, Zhou, Cai, and
  Ju}}]{qian2023anisotropic}
\bibinfo{author}{\bibfnamefont{T.-X.} \bibnamefont{Qian}},
  \bibinfo{author}{\bibfnamefont{J.}~\bibnamefont{Zhou}},
  \bibinfo{author}{\bibfnamefont{T.-Y.} \bibnamefont{Cai}}, \bibnamefont{and}
  \bibinfo{author}{\bibfnamefont{S.}~\bibnamefont{Ju}},
  \bibinfo{journal}{Physical Review Research} \textbf{\bibinfo{volume}{5}},
  \bibinfo{pages}{033143} (\bibinfo{year}{2023}).

\bibitem[{\citenamefont{Datta et~al.}(2024)\citenamefont{Datta, Adak, Yu,
  Dharmapalan, Hall, Vakulenko, Komissarenko, Kurganov, Quan, Wang
  et~al.}}]{datta2024magnon}
\bibinfo{author}{\bibfnamefont{B.}~\bibnamefont{Datta}},
  \bibinfo{author}{\bibfnamefont{P.~C.} \bibnamefont{Adak}},
  \bibinfo{author}{\bibfnamefont{S.}~\bibnamefont{Yu}},
  \bibinfo{author}{\bibfnamefont{A.~V.} \bibnamefont{Dharmapalan}},
  \bibinfo{author}{\bibfnamefont{S.~J.} \bibnamefont{Hall}},
  \bibinfo{author}{\bibfnamefont{A.}~\bibnamefont{Vakulenko}},
  \bibinfo{author}{\bibfnamefont{F.}~\bibnamefont{Komissarenko}},
  \bibinfo{author}{\bibfnamefont{E.}~\bibnamefont{Kurganov}},
  \bibinfo{author}{\bibfnamefont{J.}~\bibnamefont{Quan}},
  \bibinfo{author}{\bibfnamefont{W.}~\bibnamefont{Wang}}, \bibnamefont{et~al.},
  \bibinfo{journal}{arXiv preprint arXiv:2409.18501}  (\bibinfo{year}{2024}).

\bibitem[{\citenamefont{Moubah et~al.}(2016)\citenamefont{Moubah, Magnus,
  Warnatz, P{\'a}lsson, Kapaklis, Ukleev, Devishvili, Palisaitis, Persson, and
  Hj{\"o}rvarsson}}]{moubah2016discrete}
\bibinfo{author}{\bibfnamefont{R.}~\bibnamefont{Moubah}},
  \bibinfo{author}{\bibfnamefont{F.}~\bibnamefont{Magnus}},
  \bibinfo{author}{\bibfnamefont{T.}~\bibnamefont{Warnatz}},
  \bibinfo{author}{\bibfnamefont{G.~K.} \bibnamefont{P{\'a}lsson}},
  \bibinfo{author}{\bibfnamefont{V.}~\bibnamefont{Kapaklis}},
  \bibinfo{author}{\bibfnamefont{V.}~\bibnamefont{Ukleev}},
  \bibinfo{author}{\bibfnamefont{A.}~\bibnamefont{Devishvili}},
  \bibinfo{author}{\bibfnamefont{J.}~\bibnamefont{Palisaitis}},
  \bibinfo{author}{\bibfnamefont{P.}~\bibnamefont{Persson}}, \bibnamefont{and}
  \bibinfo{author}{\bibfnamefont{B.}~\bibnamefont{Hj{\"o}rvarsson}},
  \bibinfo{journal}{Physical Review Applied} \textbf{\bibinfo{volume}{5}},
  \bibinfo{pages}{044011} (\bibinfo{year}{2016}).

\bibitem[{\citenamefont{Marques-Moros et~al.}(2023)\citenamefont{Marques-Moros,
  Boix-Constant, Ma{\~n}as-Valero, Canet-Ferrer, and
  Coronado}}]{marques2023interplay}
\bibinfo{author}{\bibfnamefont{F.}~\bibnamefont{Marques-Moros}},
  \bibinfo{author}{\bibfnamefont{C.}~\bibnamefont{Boix-Constant}},
  \bibinfo{author}{\bibfnamefont{S.}~\bibnamefont{Ma{\~n}as-Valero}},
  \bibinfo{author}{\bibfnamefont{J.}~\bibnamefont{Canet-Ferrer}},
  \bibnamefont{and} \bibinfo{author}{\bibfnamefont{E.}~\bibnamefont{Coronado}},
  \bibinfo{journal}{ACS nano} \textbf{\bibinfo{volume}{17}},
  \bibinfo{pages}{13224} (\bibinfo{year}{2023}).

\bibitem[{\citenamefont{Pawbake et~al.}(2023)\citenamefont{Pawbake, Pelini,
  Wilson, Mosina, Sofer, Heid, and Faugeras}}]{pawbake2023raman}
\bibinfo{author}{\bibfnamefont{A.}~\bibnamefont{Pawbake}},
  \bibinfo{author}{\bibfnamefont{T.}~\bibnamefont{Pelini}},
  \bibinfo{author}{\bibfnamefont{N.~P.} \bibnamefont{Wilson}},
  \bibinfo{author}{\bibfnamefont{K.}~\bibnamefont{Mosina}},
  \bibinfo{author}{\bibfnamefont{Z.}~\bibnamefont{Sofer}},
  \bibinfo{author}{\bibfnamefont{R.}~\bibnamefont{Heid}}, \bibnamefont{and}
  \bibinfo{author}{\bibfnamefont{C.}~\bibnamefont{Faugeras}},
  \bibinfo{journal}{Physical Review B} \textbf{\bibinfo{volume}{107}},
  \bibinfo{pages}{075421} (\bibinfo{year}{2023}).

\bibitem[{\citenamefont{Klein et~al.}(2022)\citenamefont{Klein, Pham, Thomsen,
  Curtis, Denneulin, Lorke, Florian, Steinhoff, Wiscons, Luxa
  et~al.}}]{klein2022control}
\bibinfo{author}{\bibfnamefont{J.}~\bibnamefont{Klein}},
  \bibinfo{author}{\bibfnamefont{T.}~\bibnamefont{Pham}},
  \bibinfo{author}{\bibfnamefont{J.}~\bibnamefont{Thomsen}},
  \bibinfo{author}{\bibfnamefont{J.}~\bibnamefont{Curtis}},
  \bibinfo{author}{\bibfnamefont{T.}~\bibnamefont{Denneulin}},
  \bibinfo{author}{\bibfnamefont{M.}~\bibnamefont{Lorke}},
  \bibinfo{author}{\bibfnamefont{M.}~\bibnamefont{Florian}},
  \bibinfo{author}{\bibfnamefont{A.}~\bibnamefont{Steinhoff}},
  \bibinfo{author}{\bibfnamefont{R.}~\bibnamefont{Wiscons}},
  \bibinfo{author}{\bibfnamefont{J.}~\bibnamefont{Luxa}}, \bibnamefont{et~al.},
  \bibinfo{journal}{Nature Communications} \textbf{\bibinfo{volume}{13}},
  \bibinfo{pages}{5420} (\bibinfo{year}{2022}).

\bibitem[{\citenamefont{Robert et~al.}(2018)\citenamefont{Robert, Semina,
  Cadiz, Manca, Courtade, Taniguchi, Watanabe, Cai, Tongay, Lassagne
  et~al.}}]{robert2018optical}
\bibinfo{author}{\bibfnamefont{C.}~\bibnamefont{Robert}},
  \bibinfo{author}{\bibfnamefont{M.}~\bibnamefont{Semina}},
  \bibinfo{author}{\bibfnamefont{F.}~\bibnamefont{Cadiz}},
  \bibinfo{author}{\bibfnamefont{M.}~\bibnamefont{Manca}},
  \bibinfo{author}{\bibfnamefont{E.}~\bibnamefont{Courtade}},
  \bibinfo{author}{\bibfnamefont{T.}~\bibnamefont{Taniguchi}},
  \bibinfo{author}{\bibfnamefont{K.}~\bibnamefont{Watanabe}},
  \bibinfo{author}{\bibfnamefont{H.}~\bibnamefont{Cai}},
  \bibinfo{author}{\bibfnamefont{S.}~\bibnamefont{Tongay}},
  \bibinfo{author}{\bibfnamefont{B.}~\bibnamefont{Lassagne}},
  \bibnamefont{et~al.}, \bibinfo{journal}{Physical Review Materials}
  \textbf{\bibinfo{volume}{2}}, \bibinfo{pages}{011001} (\bibinfo{year}{2018}).

\bibitem[{\citenamefont{Malitson}(1965)}]{malitson1965interspecimen}
\bibinfo{author}{\bibfnamefont{I.~H.} \bibnamefont{Malitson}},
  \bibinfo{journal}{Journal of the optical society of America}
  \textbf{\bibinfo{volume}{55}}, \bibinfo{pages}{1205} (\bibinfo{year}{1965}).

\bibitem[{\citenamefont{Schinke et~al.}(2015)\citenamefont{Schinke,
  Christian~Peest, Schmidt, Brendel, Bothe, Vogt, Kr{\"o}ger, Winter,
  Schirmacher, Lim et~al.}}]{schinke2015uncertainty}
\bibinfo{author}{\bibfnamefont{C.}~\bibnamefont{Schinke}},
  \bibinfo{author}{\bibfnamefont{P.}~\bibnamefont{Christian~Peest}},
  \bibinfo{author}{\bibfnamefont{J.}~\bibnamefont{Schmidt}},
  \bibinfo{author}{\bibfnamefont{R.}~\bibnamefont{Brendel}},
  \bibinfo{author}{\bibfnamefont{K.}~\bibnamefont{Bothe}},
  \bibinfo{author}{\bibfnamefont{M.~R.} \bibnamefont{Vogt}},
  \bibinfo{author}{\bibfnamefont{I.}~\bibnamefont{Kr{\"o}ger}},
  \bibinfo{author}{\bibfnamefont{S.}~\bibnamefont{Winter}},
  \bibinfo{author}{\bibfnamefont{A.}~\bibnamefont{Schirmacher}},
  \bibinfo{author}{\bibfnamefont{S.}~\bibnamefont{Lim}}, \bibnamefont{et~al.},
  \bibinfo{journal}{Aip Advances} \textbf{\bibinfo{volume}{5}}
  (\bibinfo{year}{2015}).

\bibitem[{\citenamefont{Lee et~al.}(2019)\citenamefont{Lee, Jeong, Jung, and
  Yee}}]{lee2019refractive}
\bibinfo{author}{\bibfnamefont{S.-Y.} \bibnamefont{Lee}},
  \bibinfo{author}{\bibfnamefont{T.-Y.} \bibnamefont{Jeong}},
  \bibinfo{author}{\bibfnamefont{S.}~\bibnamefont{Jung}}, \bibnamefont{and}
  \bibinfo{author}{\bibfnamefont{K.-J.} \bibnamefont{Yee}},
  \bibinfo{journal}{physica status solidi (b)} \textbf{\bibinfo{volume}{256}},
  \bibinfo{pages}{1800417} (\bibinfo{year}{2019}).

\end{thebibliography}
\end{document}